\documentclass[a4paper,11pt]{article}
\usepackage{caption}

\usepackage{jinstpub} 
\usepackage[utf8]{inputenc}
\usepackage{lineno,hyperref,color}
\modulolinenumbers[2]
\usepackage{graphicx}
\usepackage{adjustbox}

\clearpage
\title{\boldmath CATIROC: an integrated chip for neutrino experiments using photomultiplier tubes }  

 \author[a,*]{Selma Conforti}
 \affiliation[a]{OMEGA, Ecole Polytechnique-CNRS/IN2P3, Paris, France}
 \author[b,*]{Mariangela Settimo}
 \affiliation[b]{SUBATECH, IMT Atlantique, Universit\'e de Nantes, CNRS-IN2P3, Nantes, France}

 \author[c]{Cayetano Santos}
 \affiliation[c]{Astro-Particle Physics Laboratory, CNRS/CEA/Paris7/Observatoire de Paris, Paris, France}
\author[d]{Cl\'ement Bordereau} 
\affiliation[d]{Univ. Bordeaux, CNRS, CENBG, UMR 5797, F-33170 Gradignan, France} 
\author[e,f]{Anatael Cabrera}
\affiliation[e]{IJCLab, Universit\'e Paris-Saclay, CNRS/IN2P3, 91405 Orsay, France}
\affiliation[f]{LNCA Underground Laboratory, IN2P3/CNRS - CEA, Chooz, France}
\author[a]{St\'ephane Callier}
\author[d]{C\'edric Cerna} 
\author[a]{Christophe De La Taille}
\author[d]{Fr\'ed\'eric Druillole}
\author[a]{Fr\'ed\'eric Dulucq}
\author[b]{Victor Lebrin}
 \author[b]{Fr\'ed\'eric Lef\`evre}
\author[a]{Gis\`ele Martin-Chassard}

\author[d]{Fr\'ed\'eric Perrot}
 \author[d]{Abdel Rebii} 
 \author[b]{Louis-Marie Rigalleau}
\author[a]{Nathalie Seguin-Moreau}
\author{\\on behalf of JUNO collaboration}
\emailAdd{conforti@omega.in2p3.fr}
\emailAdd{settimo@subatech.in2p3.fr}

 \abstract{
     An ASIC (Application Specific Integrated Chip) named CATIROC (Charge And Time Integrated Read Out Chip) has been developed for the next-generation neutrino experiments using a large number of photomultiplier tubes (PMTs). Each CATIROC provides the time and the charge measurements for 16 configurable input channels operating in auto-trigger mode. 
    Originally designed for the light emission  in water Cherenkov detectors, we show in this paper that its use can be extended to liquid-scintillator based experiments. The $\sim$~26000 3-inch PMTs of the JUNO experiment, under construction in China, is a case in point.  This paper describes the features of  CATIROC with a special attention to the most critical points for its application to the time profile of the light emission in liquid scintillators. The achieved performances in both charge and time measurements can be inputs for future high-precision experiments making use of PMTs or other photo-sensitive detectors. 
}

\begin{document}
\maketitle
\newpage

\section{Introduction} \label{sect:Intro}

Even though neutrinos were discovered over half a century ago, neutrino physics is still one of the most active fields in physics. A precise measurement of the neutrino oscillations and of the  neutrino mass ordering is requiring a huge exposure and challenging performances for the new experiments. 
Multi-tons experiments using thousands of photomultiplier tubes (PMTs) are currently under design and construction (e.g., JUNO~\cite{JUNO-physics,JUNO-concept}, HyperK~\cite{HyperK}, KM3NeT-ORCA~\cite{orca},...). For such experiments, flexible integrated systems with fast response,  high performance in charge and time determination and compact data stream become more and more crucial. The typical requirements are listed in Table~\ref{tab:req}. 

The Jiangmen Underground Neutrino Observatory (JUNO), currently under construction in the south of China, aims to determine the neutrino mass ordering, namely the sign of the atmospheric mass splitting and to perform precision measurement of the oscillation parameters. With its 20~kilotons of liquid scintillator (LS), it will be the biggest detector of such a type ever built, and it will be competitive for astrophysical neutrinos, as Supernovae, geo-netrinos and for rare-events physics.   
The detector consists of a spherical array of 18000 20-inch PMTs and about 26000 3-inch PMTs (also named ``small PMT'' or ``SPMT''). The core of the SPMTs readout is the CATIROC chip,  a 16-channel front-end ASIC (Application Specific Integrated Chip) designed by the Omega laboratory in AustriaMicroSystems (AMS) SiGe 0.35 $\mu$m technology to readout PMTs in large-scale applications. CATIROC is an upgraded version of PARISROC2~\cite{parisroc} chip conceived in 2010 in the context of the R\&D PMm$^{2}$ (square meter PhotoMultiplier) project~\cite{PMm2}. 

CATIROC, was initially developed for large water Cherenkov detectors. Thus, its application in liquid-scintillator based experiments requires some additional considerations. In fact, differently from the Cherenkov light emission which arrives to the PMTs in a few nanoseconds, the processes of excitations and re-emission of photons in liquid scintillator determine a signal spread on longer  time windows  (up to hundreds of nanoseconds depending on the LS specific recipe). The time separation between consecutive hits in the same PMT is thus significantly large and the matching between the dead time and the signal time distribution becomes critical. 

In this paper we point out some new features of CATIROC that were not examined in previous characterization of the chip, as in~\cite{catiroc} or in the datasheet.  We perform dedicated measurements to ensure that the detection of multiple hits, which arrive with a time separation  typical of the LS, (from tens of ns up to hundreds of ns), is not biased. A dead time on the time scale of tens of ns is emerged from these measurements. Moreover, we extensively study the charge pile-up (or charge acceptance) in the case of two hits arriving shortly in time, especially close to this trigger dead time window. We prove that some signal loss or underestimation will occur if the CATIROC configuration is not properly adapted to the physics of the detector. Several  CATIROC configurable parameters have been tested to find the optimal ones which mitigate the charge biases and trigger dead time effects.  Finally, we show that the application of CATIROC to a liquid scintillator is actually possible only when the rate of multiples hits on the same PMT is sub-dominant. This is the case of  PMTs  operating in ``photon-counting" mode  as, for example, for the primary physics goal of the JUNO SPMT sub-system~\cite{spmtMiao, SPMT-paper}. 
Because of the small photocathode surface, each 3-inch PMT will rarely detect more than one photo-electron (PE) from neutrino interaction events and will thus suit the CATIROC features. For 1 MeV positron uniformly distributed in the detector, about 1\% of the SPMTs have more than one photon hitting the PMT photocathode (4\% at 10 MeV). For them, the time separation between hits in the same PMT is within 250 ns in 99\% of the cases. 
For completeness, we also provide a description of the major features of the CATIROC, its charge and time response, its trigger efficiency and the performance attainable even in the cases of PMTs with multiple-hits (i.e., atmospheric muons crossing the detector). 

\vspace{0.3cm}
\begin{tabular}{lcc}
\hline\hline
Requirement  & value & section \\
\hline\hline
Number of channels & O(10$^4$) & \ref{sect:Intro} \\
Trigger Threshold & $<$ 1/3 PE & \ref{sect:discriminator} \\ 
Trigger mode & auto-trigger  & \ref{sect:discriminator} \\ 
Dead time & application specific (O(10~$\rm{\mu}$s)) & \ref{sect:deadtime} \\ 
Dynamic range & application specific (1 to $\sim$ 100 PE) &  \ref{sect:charge}  \\
Charge resolution & $<$ 0.1 PE &  \ref{sect:charge} \\
Charge linearity deviation & $<$ 1\% & \ref{sect:charge} \\ 
Cross talk & O(10$^{-3}$) &  \ref{sect:charge} \\
Charge integration window & application specific ($\sim$20~ns) & \ref{sect:shaper} \\
Time resolution &  O(100 ps) & \ref{sect:time} \\
\hline\hline
\end{tabular}
\captionof{table}{Typical requirements for a PMT-based experiment. Some of the requirements are related to the physics goals (e.g., the expected signal intensity and variation for the charge dynamic range, the expected trigger rate for the dead time, the time profile for the time accuracy), the type of experiment (Cherenkov or liquid scintillator which affects the time integration and dead time) or the detector performance (e.g., detector size and material, PMT signal width and transit time spread, trigger conditions). For these cases we indicate between parentheses the expected values for the JUNO-SPMT application.\\}
\label{tab:req}
\vspace{0.3cm}

The manuscript is organized as follows. The CATIROC design is described in Section~\ref{sect:catiroc}. The main tests to characterize the ASIC performance are presented in Section~\ref{sect:measurements} which  includes the trigger efficiency, the dead time contributions, the charge linearity and resolution and the time resolution.  For each specific performance test, the experimental requirements are discussed in the related sub-section. In Section~\ref{sect:spmt} we performed tests of CATIROC with 3-inch PMTs. The suitability of CATIROC to fulfill the requirements of liquid scintillator experiments is finally summarized in Section~\ref{sect:discussions}. 

\begin{figure}[b!]
  \includegraphics[width=\textwidth]{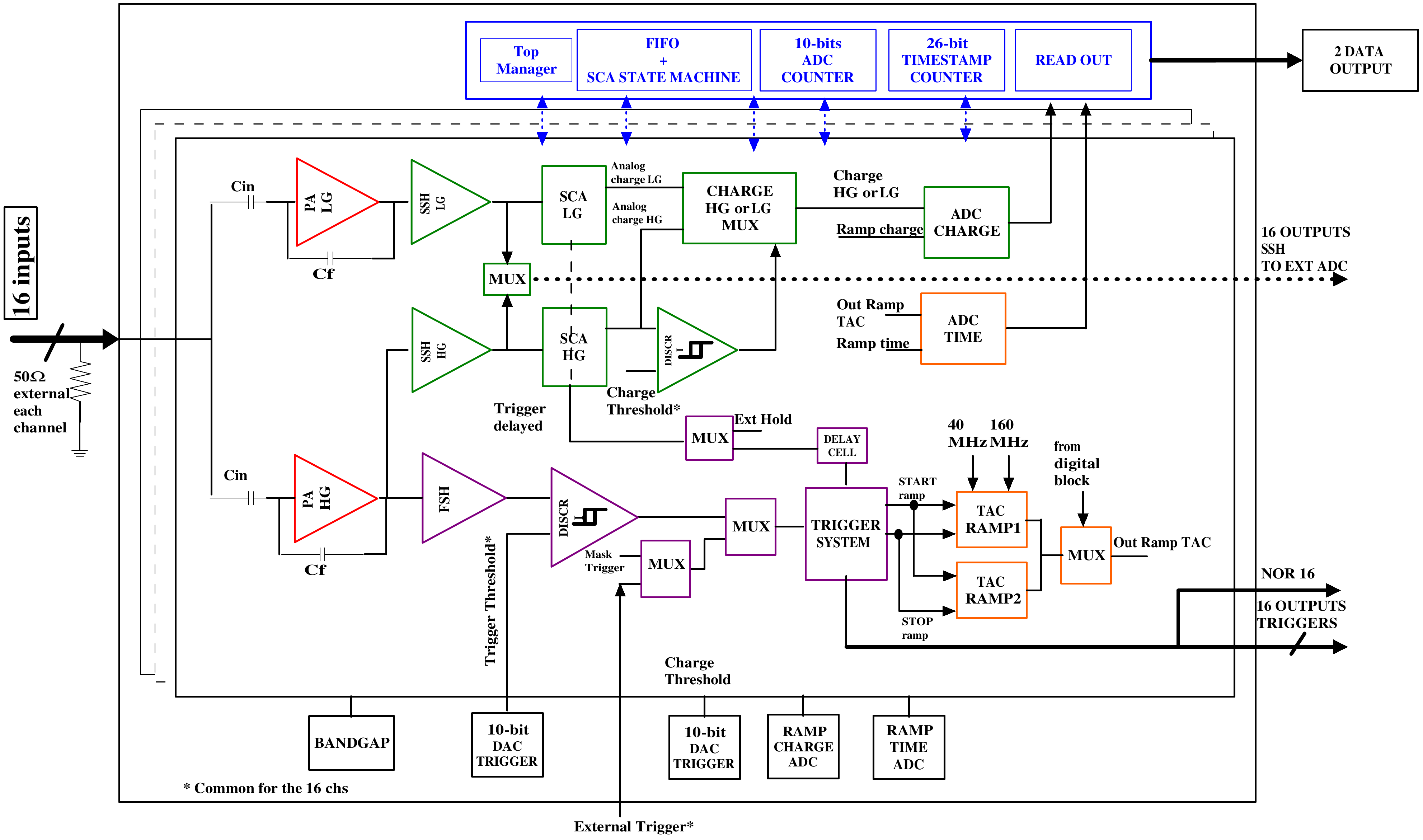}\\
  \includegraphics[width=0.9\textwidth]{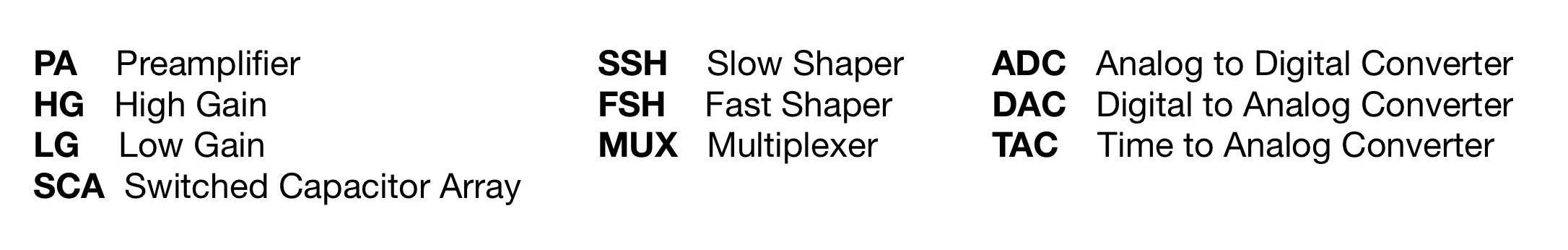}
  \caption{CATIROC simplified block schematic. In violet, the part responsible for the trigger formation; in orange and green the part dedicated to time and integrated-charge measurements and in red the two preamplifiers; in blue and black the digital parts of the circuit and the block commons to all channels.   }
  \label{fig:design}
\end{figure}

\section{The CATIROC design} \label{sect:catiroc} 

CATIROC is conceived as an autonomous and flexible system made of a chip operating in auto-trigger mode and providing a measurement of the arrival time and the integrated-charge for 16 input signals. An adjustable gain is featured for each input channel to compensate for the gain variation between the PMTs. A shift register is used to send the configuration parameters (hereafter named slow control parameters) inside the chip. There are 328 slow control parameters which are loaded serially to control the chip. The architecture of the ASIC is shown in Fig.~\ref{fig:design}, distinguishing in violet, orange and green the circuits responsible for the trigger operation, and time and charge measurements, respectively. 
In the following we describe these circuital parts with a level of details sufficient for a comprehensive view of the chip functionalities and performance.  Other technical informations are given in the CATIROC data sheet which can be provided on request. \\ 
Each of the 16 input channels has two low-noise preamplifiers (PA), for the high (HG) and the low gain (LG) paths with a fixed ratio LG/HG of 10. The two preamplifiers have an input capacitance (C$_{\rm{in}}$) of  5 pF and 0.5 pF, respectively. A feedback variable capacitor ($C_f$) tunable with a 8 bits slow control parameter,  is used to further vary the gain of each channel independently (see Figure~\ref{fig:design}). The gain is obtained from the ratio C$_{\rm{in}}$/C$_f$ and can theoretically vary in a 8 bits range with the smallest step of 0.008~pF and within the limit of the signal saturation (see Section~\ref{sect:charge}). The use of the HG and LG paths allows to achieve a dynamic range from 160 fC up to 70 pC (from 0.3 PE to145 PEs  at a PMT gain of $3 \times 10^6$ and with a preamplifier gain of 20).   After the preamplification phase, the signal feeds a slow and a fast channels used respectively for a charge measurement and for timing  and trigger. 

\paragraph{\bf Fast channel}
It is responsible for the trigger formation and the time measurement (violet and orange circuital parts in Fig.~\ref{fig:design}). It comprises a fast shaper (FSH), followed by a low-offset discriminator which allows to auto-trigger on signals above a set threshold. \\
The fast shaper is a band-pass (CR/RC) circuit with a time constant of 5~ns. It filters possible noise contributions that would also alter the minimum signal threshold. It is coupled to the HG preamplifier to produce a trigger on input charges as low as 50 fC.  For each channel, a logical gate system is used to select between the trigger formed by the input signal and an external trigger (injected through an external system as for example an FPGA). In addition, an internal structure allows to mask the internal trigger in each channel. 

The discriminator output is followed by a system responsible for the time measurement and for storing the analog signal charge of the ``Slow channel'' discuss below. 
The time is composed of two terms: 
\begin{equation}
T_{\rm{tot}}~\rm{[ns]}= \rm{CoarseTime} \times \rm{~25~[ns]~-~FineTime~[ns].}
\end{equation}
where the ``coarse time'' results from a 26-bit Gray Counter with a resolution of 25 ns (40 MHz clock) 
and the ``fine time''  uses a Time to Analog Converter (TAC) and a 10-bit ADC to provide a time measurement inside the 25 ns coarse time clock with a resolution (in RMS) of less than 200~ps.  

\begin{figure}[t!]
\centering
 \includegraphics[width=0.7\textwidth]{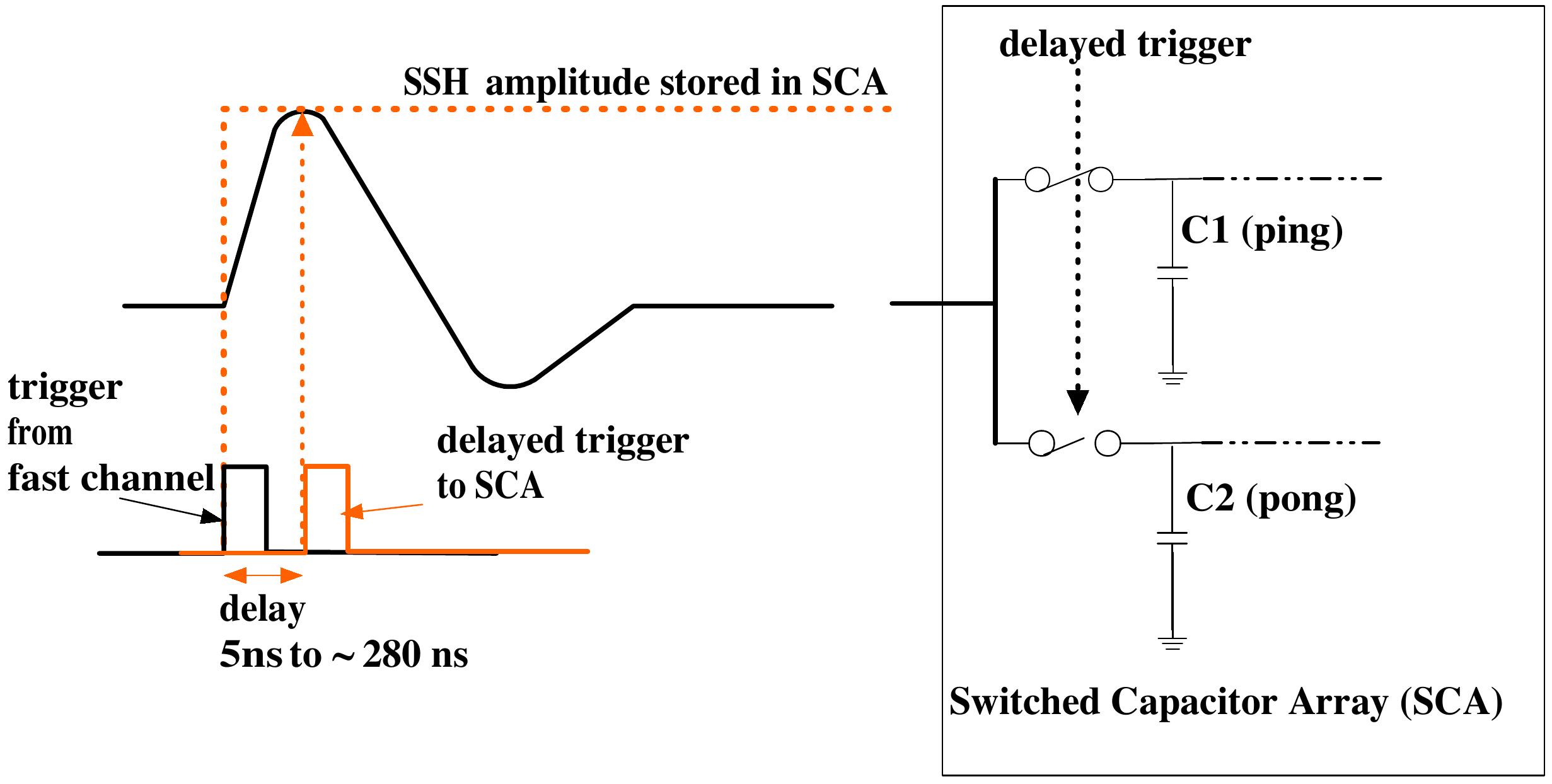}
  \caption{The Switched Capacitor Array (SCA) schematic. The trigger signal output is delayed by a user-configured time to match the maximum SSH voltage value. This value is stored in the first capacitor of the SCA (C1 or ping) and then digitized. The next signal follows the same path but will be stored in the second capacitor (C2 or pong).  }
  \label{fig:SCA}
\end{figure}

\paragraph{\bf Slow channel}
This part of the circuit (green in the figure) is responsible for the charge measurement and consists of a slow shaper (SSH) and a Switched Capacitor Array (SCA) system. The SCA is made of two capacitors working in the so-called ``ping-pong mode":  the analog signal is held  alternatively in two capacitors before the digitization phase, effectively reducing the dead time of the circuit. Small differences between the two capacitors will reflect on the measured charge value, but they are constant and can be fully characterized as discussed later on in the paper. 
The charge measurement process is sketched in Fig.~\ref{fig:SCA} to help the reader understanding some crucial aspects for the application of CATIROC to liquid scintillator experiments (Section~\ref{sect:shaper}). 
The circuit is repeated twice, for the HG and the LG channels, but only one of these two analog signals is selected and digitized  by an internal 10-bit Wilkinson ADC. The choice of the HG or LG charge is done by a discriminator that compares the HG SSH signal with a threshold (named ``charge threshold'') set by an internal 10-bit DAC and common to the 16 inputs. \\

All the channels are handled independently by the digital part, and only those channels that fired a trigger are digitized, transferred to the internal memory and then sent out in a data-driven way. When the charge and the time are converted and stored in the registers, the digital part stops the conversion and starts the readout (RO) of the data. The coded data are sent in parallel in two 8 channels serial links to be readout. 
The $T_{\rm{DeadTime}}$ thus is given by the convertion and readout times and is estimated up to 9.3~$\mu s$ if 8 channels per link are hit and around 6.8~$\mu s$ if only one channel is hit.  

 A specific input line (named ``incalib'')  allows the injection of the input charge in more than one channel in parallel. This is possible thanks to an array of 16 switches managed by slow control. 
Finally, in addition to these main features, other outputs are accessible, as the 16 triggers, the analog signals (slow and fast shapers, preamplifiers), the digital ones. These outputs are meant for debugging but a possible application of the trigger analog signal to reduce the effective dead time will be discussed in the next section. 

\begin{figure}[tbh]
  \hspace{1cm}
  \begin{center}
   \hspace{+0.6cm}
    \includegraphics[width=0.7\textwidth]{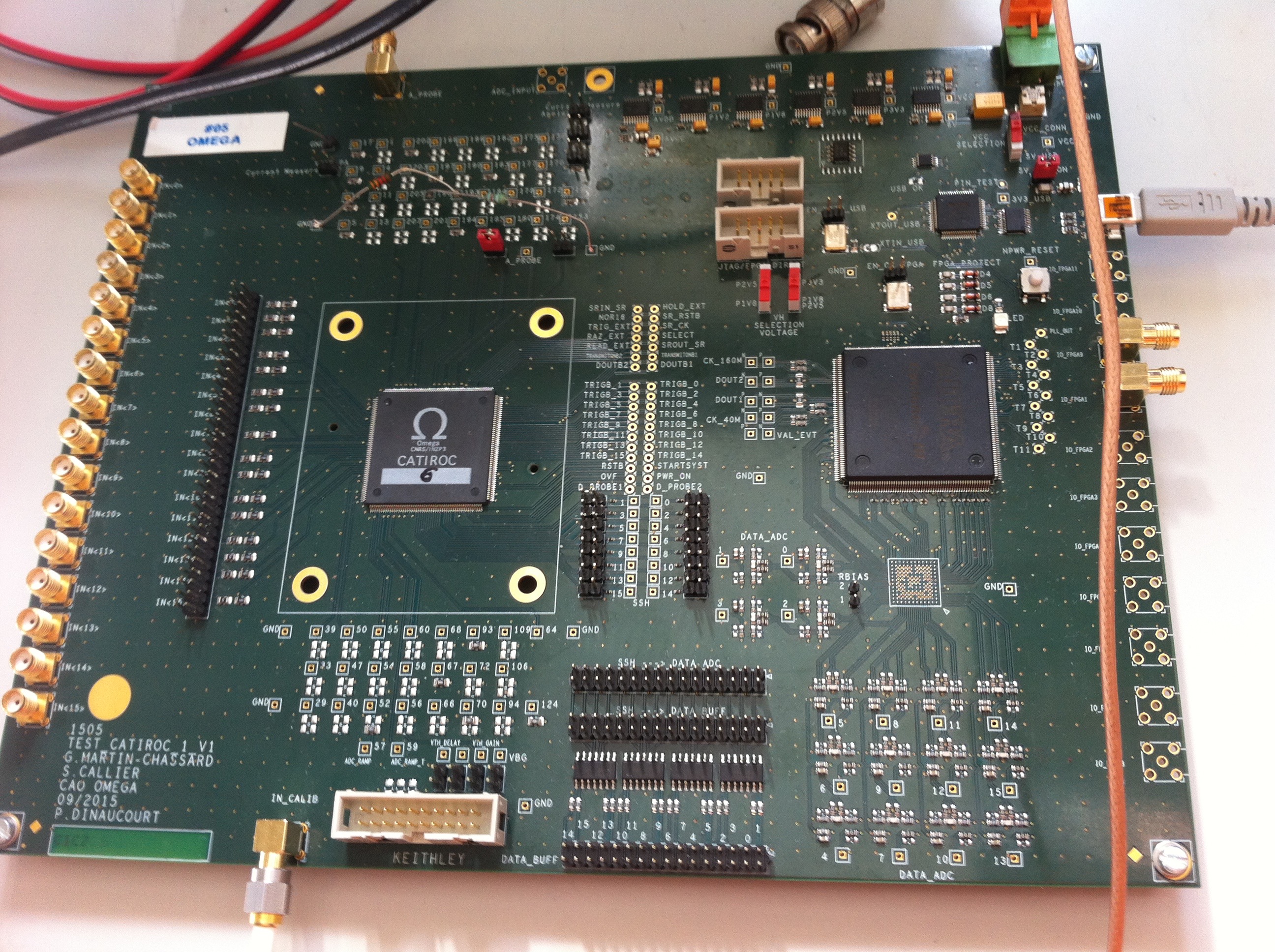}\\
    \includegraphics[width=0.8\textwidth]{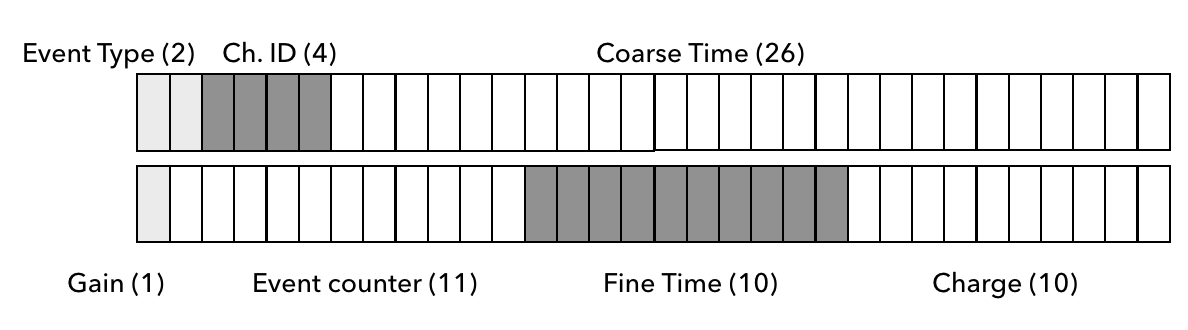}
  \end{center}
  \caption{Top: Evaluation board of CATIROC. It makes use of a general purpose communication interface (USB), for data transferring to a computer, coupled to a programmable logic array (FPGA)  Altera Cyclone III for the  implementation of custom functionalities. Bottom: the data output format used in this paper. In addition to the standard CATIROC output (3 bits for channel number, 26 bits for coarse time, 10 bits for the fine time, 10 bits for the analog charge and 1 bit for the gain), two bits are used to define the event type (standard CATIROC output, a trigger rate monitor and a debugging output), 11 bits for an event counter and an additional bit for the channel number for future uses on multi-chip boards. }
  \label{fig:testboard}
\end{figure}

\begin{figure}[tbh]
  \begin{center}
   \hspace{+0.6cm}
    \includegraphics[width=0.6\textwidth]{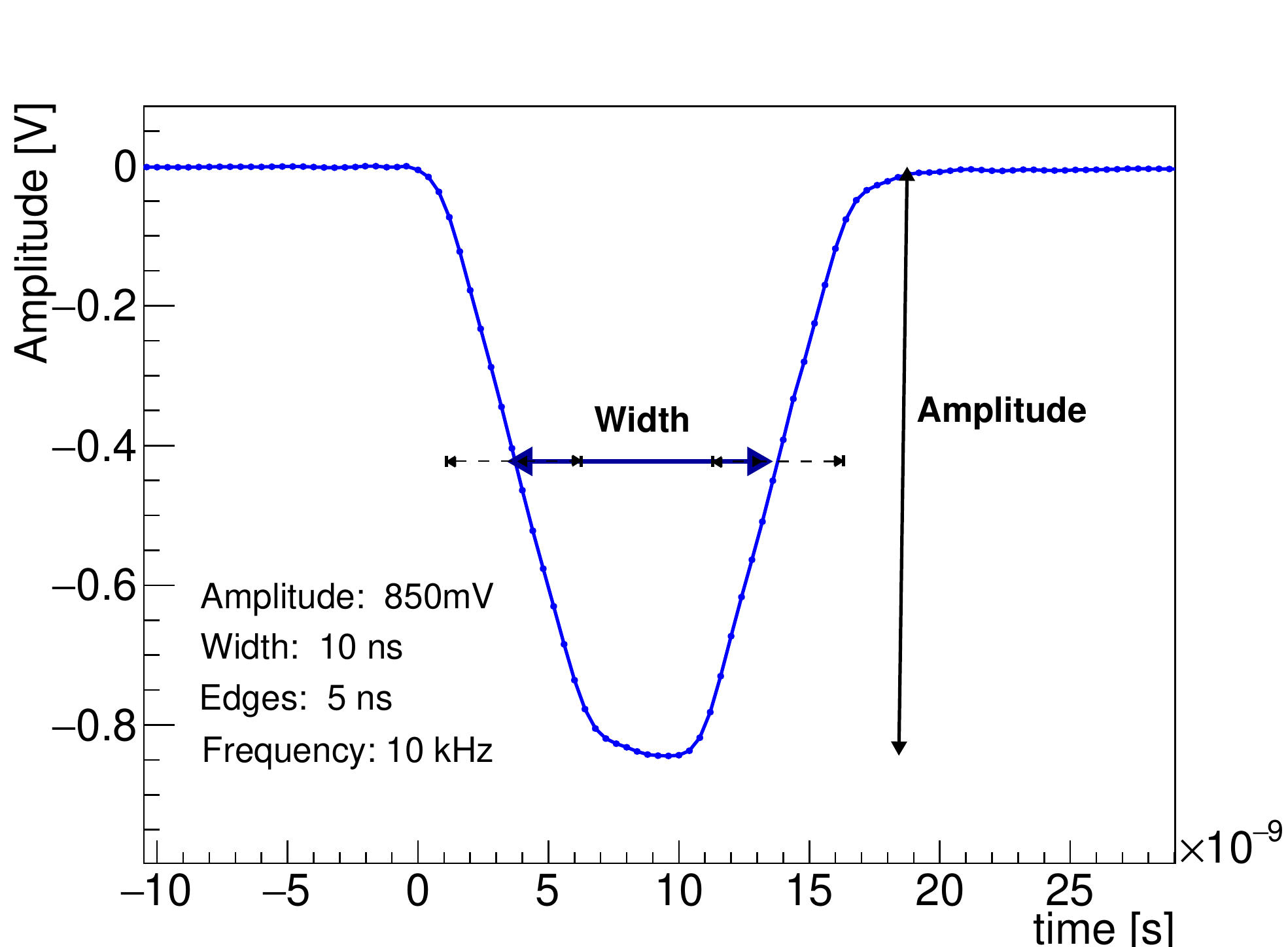}
  \end{center}
  \caption{Example of the typical pulse injected with the signal generator, having  a width of 10~ns and the edge times (calculated between 10\% and 90\% of the signal) of 5~ns. The amplitude, the edge times, frequency  and signal width have been varied depending on the performed tests. }
  \label{fig:inputsignal}
\end{figure}

\section{Test setup}\label{sect:measurements}

The evaluation board used to characterize the CATIROC ASIC is shown in Fig.~\ref{fig:testboard} (top). 
It gives access to all the 16 input channels, to the ``incalib'' line  and all CATIROC probes to monitor the analog processing of the input signal and to inspect the behavior of the digital ASIC section.
A programmable logic array (FPGA) handles the communication with the control software, sets up of the ASIC parameters and provides the two 40 and 160 MHz clocks used by CATIROC ensuring local synchronisation in multiple-chip read-out boards.  Moreover the FPGA reads out the ASIC output providing a package data 
customized as shown in Fig.~\ref{fig:testboard} (bottom).

\begin{figure}[t!]
  \hspace{-0.4cm}
  \includegraphics[width=0.56\textwidth]{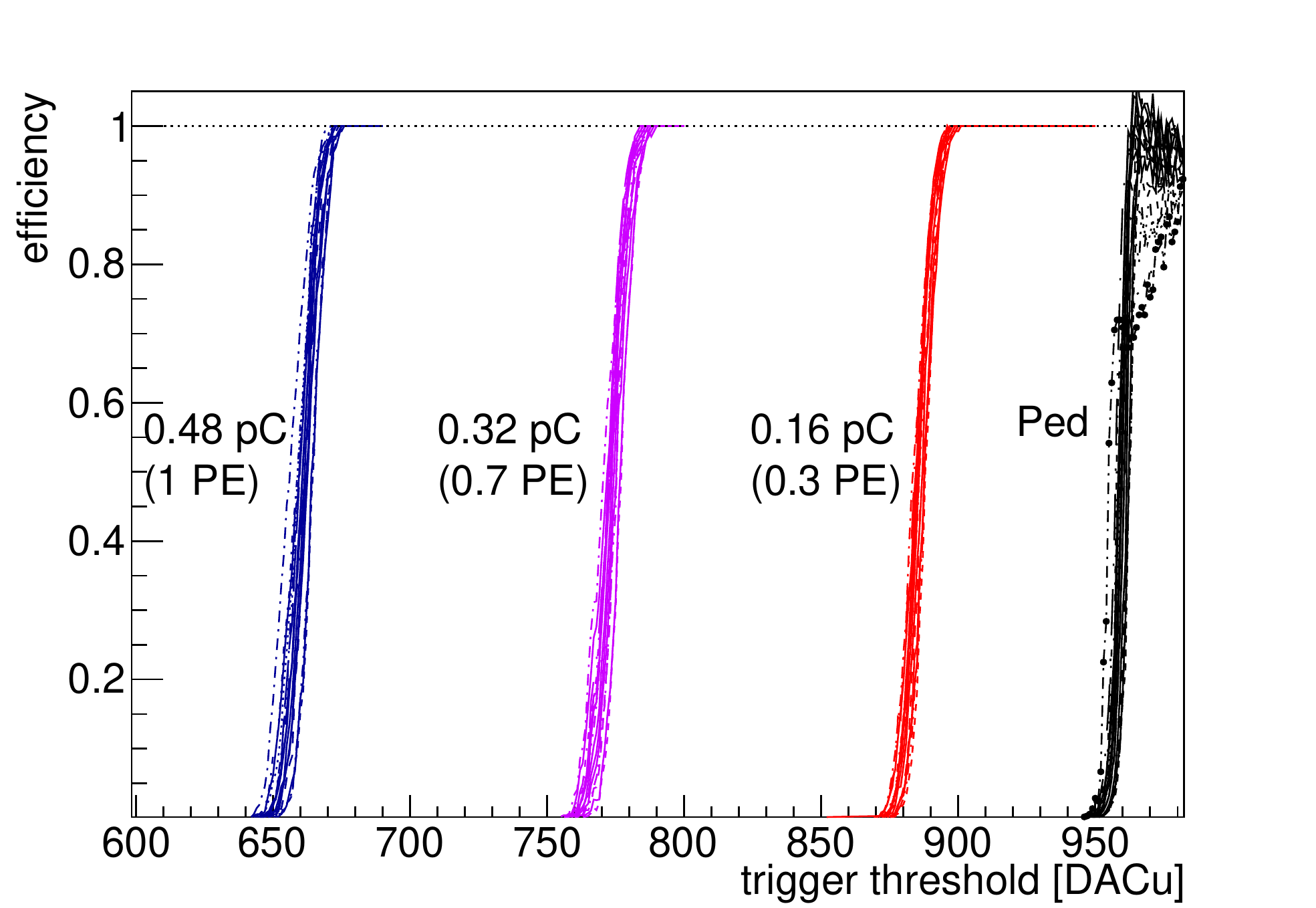}
  \hspace{-0.8cm}
   \includegraphics[width=0.56\textwidth]{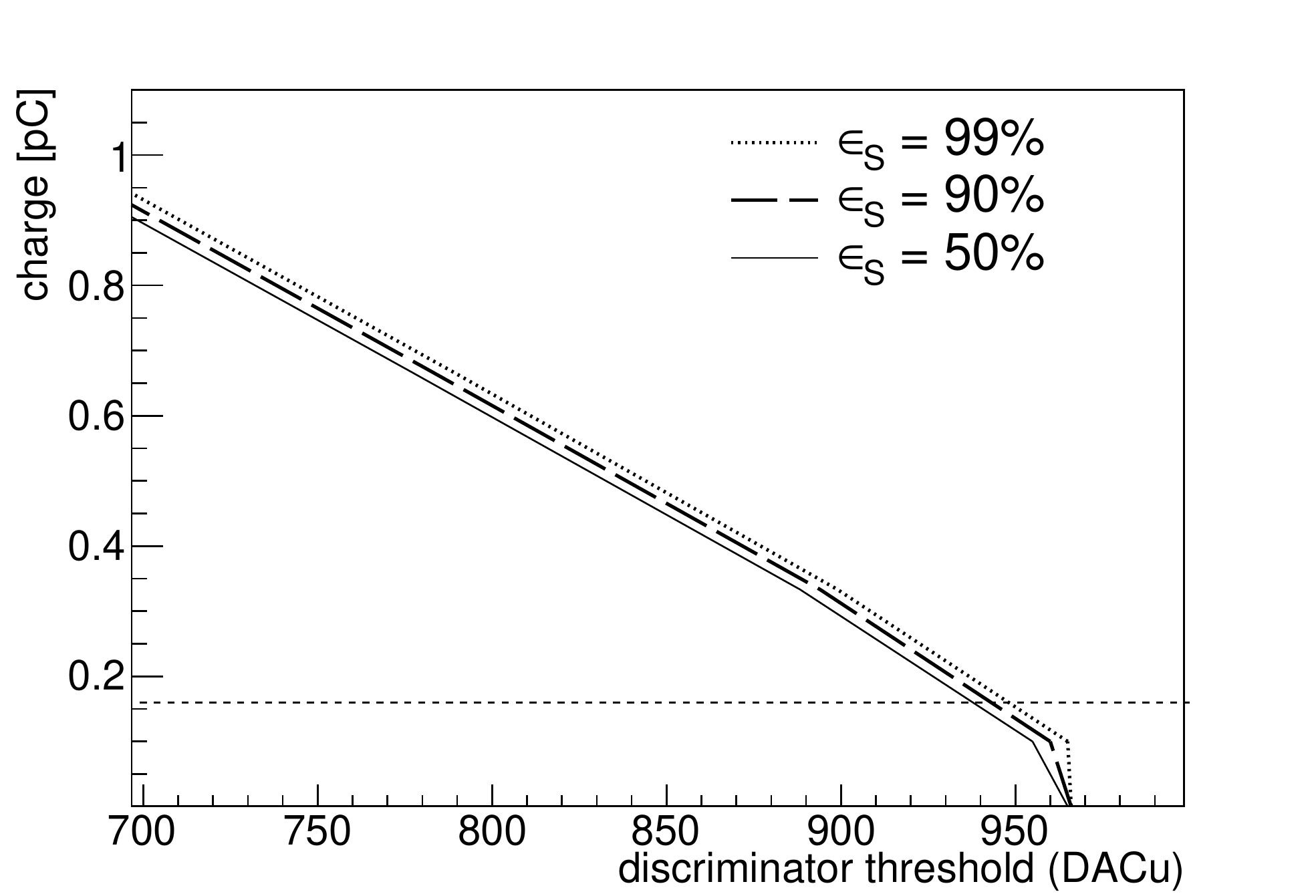}
  \caption{Left: Trigger efficiency as a function of the threshold for various injected charges up to 0.48 pC and a preamplifier gain of 20. The pedestal is on the right side and the larger inputs go to the left side of the figure because the fast shaper provides negative signals with a pedestal at around 1.8V~\cite{catiroc}. The corresponding signal value in PE, is also indicated, considering a gain of $3 \times 10^6$. Right: typical charges that can be detected with a trigger efficiency of 99\% (dotted), 90\% (long dashed) and 50\% (solid line) for a given discriminator threshold. 
  The case of 0.3 PE (i.e., 0.16~pC at a gain of $3 \times 10^6$) is shown as short dashed line for illustrative purposes.}
  \label{fig:S-curve}
\end{figure}

The tests shown in the next sub-sections are performed using a signal generator which provides pulse waveforms with edge times  $\leq$~2.5~ns, a  width of 4~ns, a  jitter smaller than 35~ps and varying the amplitudes in the range of 10 mV-10 V with signal frequencies up to 1~MHz.  A pulse of 10~ns width and 5~ns edge time is injected into an input channel to mimic a signal from a PMT. An example of the injected pulse is shown in Fig.~\ref{fig:inputsignal}. In Section~\ref{sect:shaper},  pulses with a width of 4~ns are also used to simulate consecutive hits occurring very shortly in time (within tens of ns). The amplitude is adjusted in order to scan a range of charges from a few fC up to about 60~pC. The pulse frequency, typically set at 10 kHz, is varied up to 1 MHz for dead time studies. Unless differently specified, the ASIC preamplifiers gain is set to 20 on all channels and the ``charge threshold'' is at 720 DACu, which is equivalent to a peak value of 776 mV of the signal shaper for an injected charge of about 7 pC (or 15 PE for a PMT with a gain of $3 \times 10^6$) with the current configuration. For the studies in this paper a few other parameters in the slow control (as the discriminator threshold) have been varied depending on the performed test and will be explicitly indicated in the text when needed. \\

In the following, we summarize the tests performed with CATIROC with a focus on the ones interesting for the application to scintillation light emission. More specifically, these tests include:  the trigger efficiency versus the discriminator threshold to verify the capability of observing a fraction of PE; the charge resolution and linearity over a wide signal range; the characterization of the charge biases induced by the circuit for signals arriving very close in time; the time resolution and possible dead times. 

\section{Trigger efficiency}\label{sect:discriminator}

A typical requirement of PMT-based experiments is the capability of measuring signals whose amplitude can also correspond to a fraction of a photo-electron with a dead time small enough not to affect our capabilities of reconstructing the event of interest for our physics goal. 

The trigger efficiency is defined as the ratio between the number of trigger counts in the output of CATIROC (``trigger output'') and the number of injected pulses for a given threshold. The efficiency versus the discriminator threshold (also called S-curve) is shown in Fig.~\ref{fig:S-curve} (left) for different input signals (0.16 pC, 0.32 pC and 0.48 pC). Each thin line is one of the 16 input channels. The black lines refer to the case in which no signal is injected and the trigger is fired on the pedestal. The value of the discriminator threshold is configured by a 10-bit Digital-to-Analog Converter (DAC), covering a range between 1 and 1.9 V, with a slope of 0.9 mV/DACu (${\rm Threshold [mV]} = 1000 + 0.9 \times {\rm DACu}$).  The right plot shows a projection of the S-curves and gives the minimum charge corresponding to three fixed reference values of signal efficiency (99\%,  90\% and 50\%) as a function of the trigger threshold. Any signal above these three lines will have a trigger efficiency larger than the reference values. 
A threshold of 0.3 PE (equivalent  to 0.16 pC for the case considered in the figure)  is typically considered in experiment using PMTs and is thus attainable with CATIROC. 
However, the discriminator threshold is common to all channels. This implies that an optimal value needs to be tuned for each CATIROC during the readout boards and PMTs configuration.  

\begin{figure}[tb]
\centering
     \includegraphics[width=0.7\textwidth]{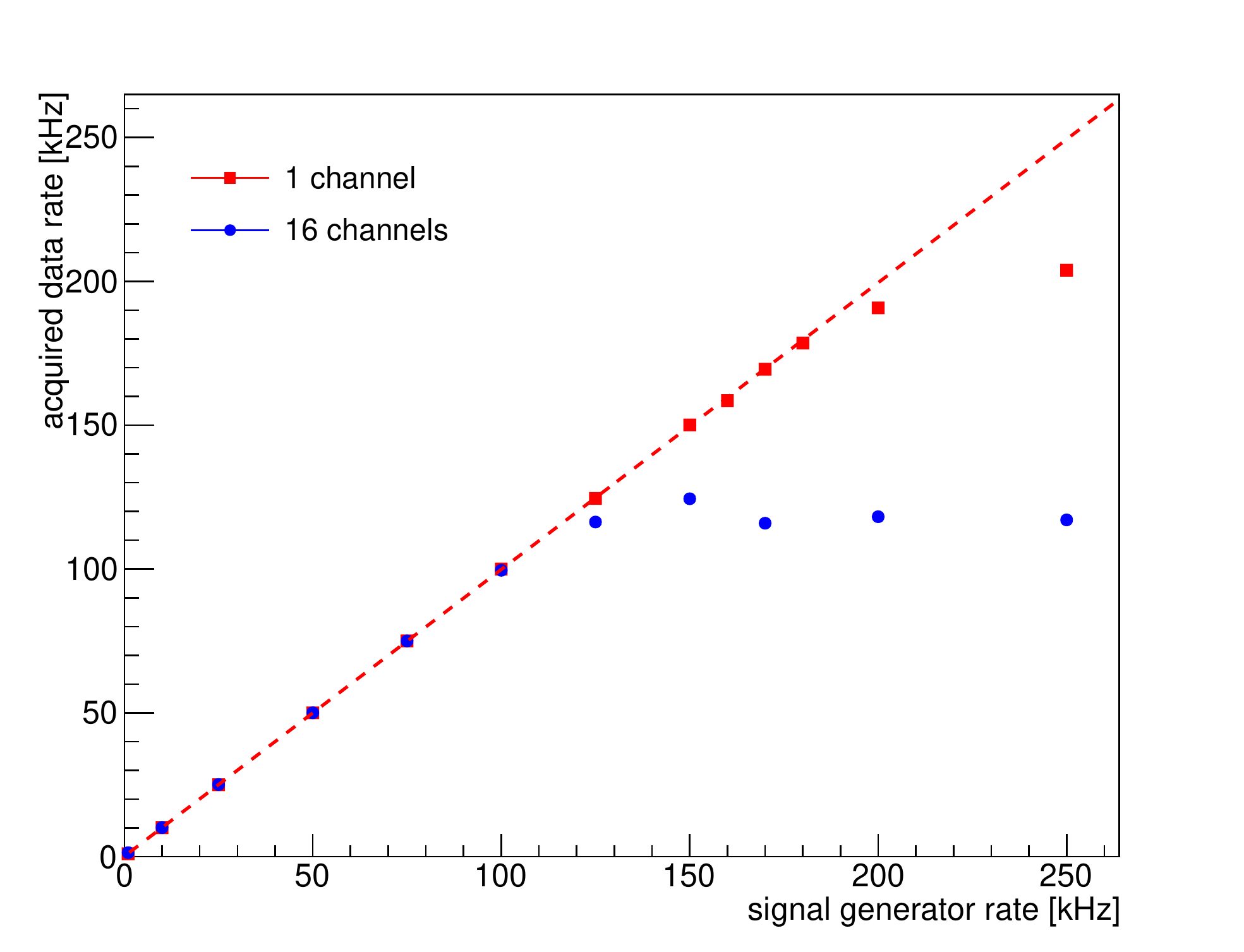}
    \caption{Trigger rate as measured with CATIROC versus the signal generator frequency, with red squares for the case of one channel hit and blue dots for the case of all channels hit at the same time}\label{fig:rate}
     \end{figure}

\section{Dead time} \label{sect:deadtime}

Two dead time contributions have been identified which may have an impact for CATIROC application to liquid scintillator-based experiments.

A first term, on a few microseconds scale, is the $T_{\rm{DeadTime}}$ already described in Section~\ref{sect:catiroc} which is due to the digitization of the charge and of the fine time. 
This dead time $T_{\rm{DeadTime}}$ has been calculated up to 6.8~$\mu$s and 9.3~$\mu$s depending on the number of data streams to read out (see Section~\ref{sect:catiroc}) and it is mitigated further by the use of a SCA with two layers of charge holding (ping and pong). This implies that up to two signals can be read out without losses if they arrive within $T_{\rm{DeadTime}}$. A third signal in the same window would be lost in the data stream.  A check of this dead time has been performed by measuring the rate of events in the CATIROC data output for increasing input signal frequencies.  Fig.~\ref{fig:rate} shows the rate of events observed with CATIROC as a function of the generator frequency for the case of signals injected in one or all channels. A saturation is observed at generator frequencies of $\sim$~180~kHz and 120~kHz for the two cases respectively, in agreement with expected values.

 \begin{figure}[tb]
  \centering
  \includegraphics[width=0.7\textwidth]{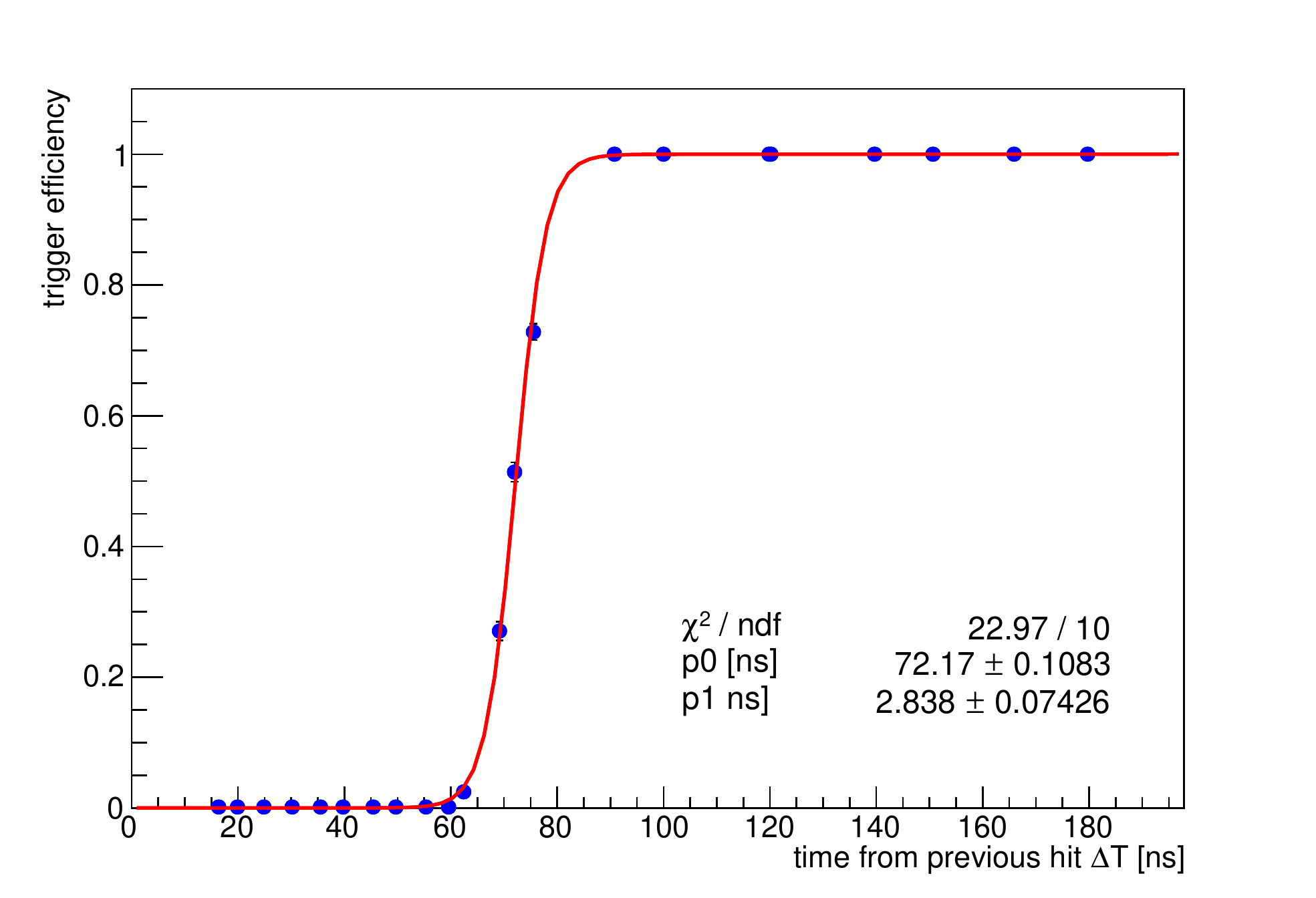}
    \caption{Trigger efficiency as a function of the time separation between two consecutive hits : an hit arriving within about 60~ns from the previous one, will not produce any independent trigger signal and the associated charge may sum up to the previous hit.   }
  \label{fig:DeadTime}
\end{figure}

\begin{figure}[t!]
  \includegraphics[width=0.48\textwidth]{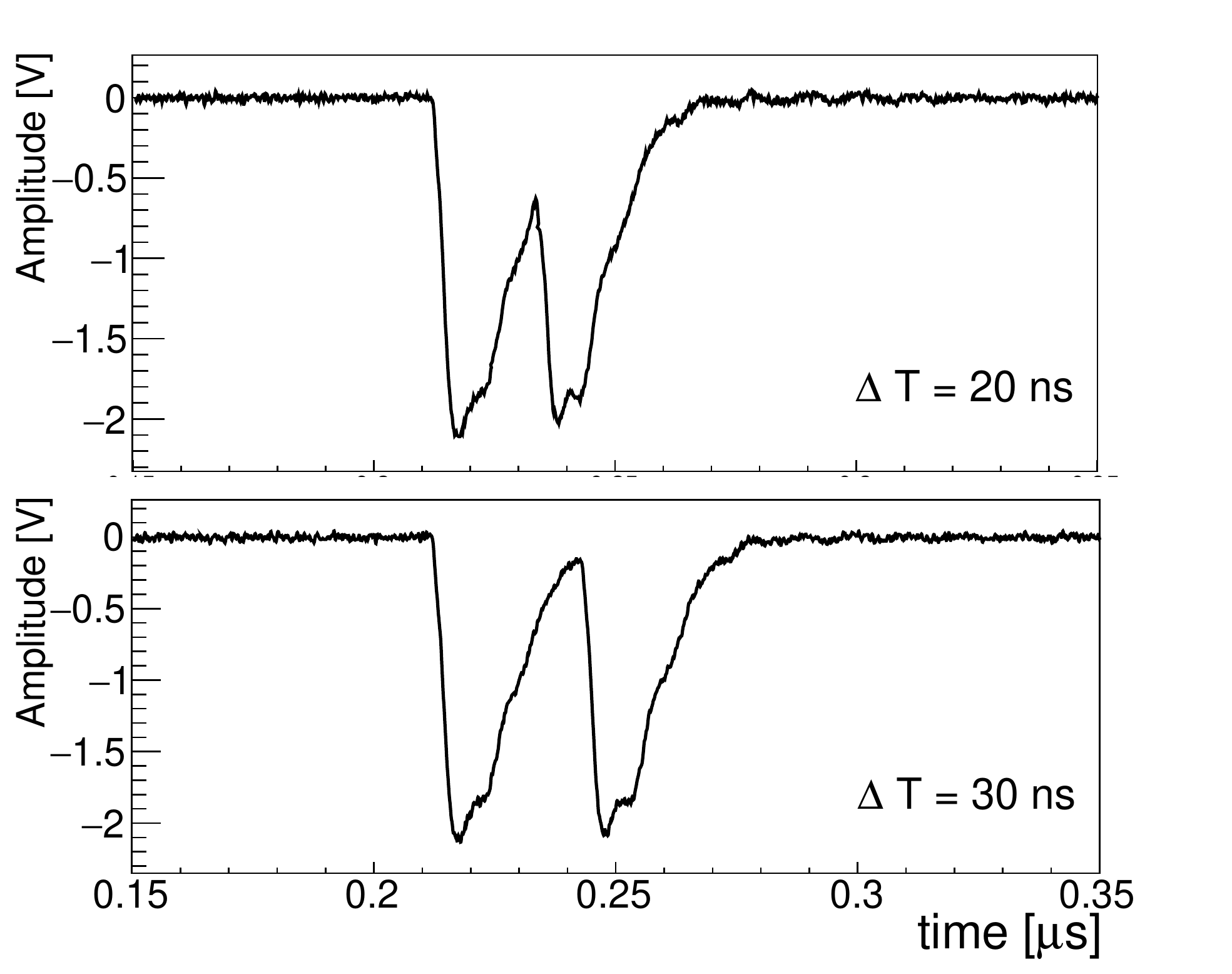}
  \hspace{-0.3cm}
   \includegraphics[width=0.59\textwidth]{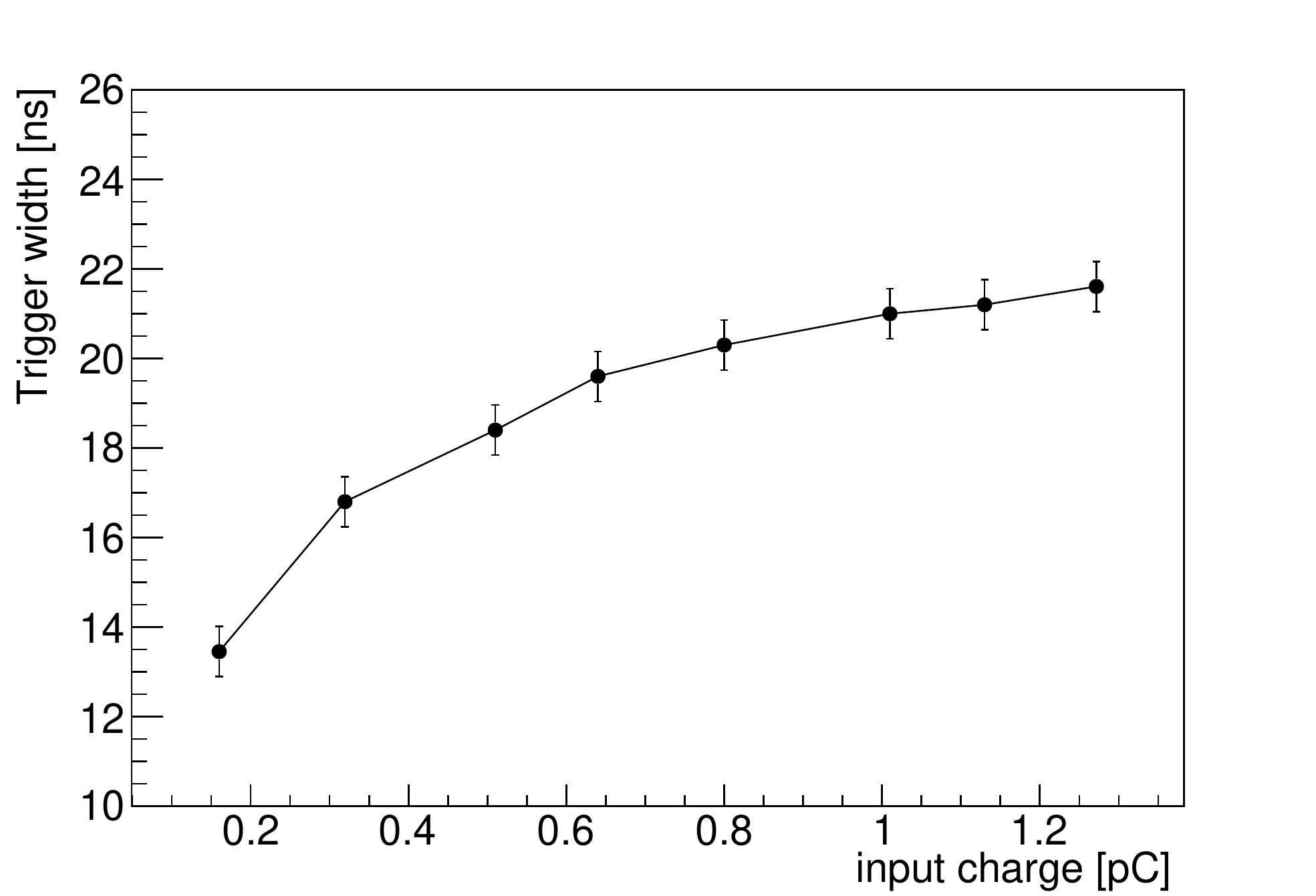}
    \caption{Left:  discriminator analog signal as observed with an oscilloscope for two pulses injected with a time difference of 20 ns (top) and 30~ns (bottom), smaller than the  $T_{\rm{TrigDeadTime}}$.  Right: the width of each trigger signal increases with the input charge up to 1~pC (about 2 PE for a gain of $3 \times 10^6$). }
  \label{fig:discri}
\end{figure}

The second dead time contributor, hereafter called $T_{\rm{TrigDeadTime}}$, is due to the handling of the trigger which is delayed to store the analog charge in the analog memory and is used to create the TAC ramp. To measure this dead time a burst of two consecutive pulses is injected with a  period $P = 1$~ms and a time separation between the two pulses ($\Delta T$) varying between 15 ns and hundreds of ns.  The obtained efficiency curve is shown in Fig.~\ref{fig:DeadTime}. A sigmoid function $1 - (1 + \exp(x - p0)/p1)^{-1})$ is fit to data. The best-fit parameter p0 indicates the time $\Delta t$, which corresponds to 50\% trigger efficiency. The p1 parameters is the inverse of the slope of the curve :  the value $p0 \pm p1$ will delimit the interval with trigger probability between 23\% and 73\%.  This curve has the same behavior in the three circuits tested for this study. Two pulses arriving with $\Delta T$ larger than 90~ns will always be detected as separate signals and the charge will be measured in the digital part of CATIROC. This is not always the case below 90~ns. For a  $\Delta T \leq 60$~ns, the second pulse is not handled by the digital part of the trigger and in practice the board does not produce an independent charge measurement in the CATIROC output. As we will discuss in Section~\ref{sect:shaper}, in this case, the charge of the second pulse sums up to the first hit, depending on their relative time difference, so that the information is not completely lost (charge acceptance). Moreover, as mentioned in the previous section, the direct signal from the discriminator, which is not affected by the digital treatment of the trigger, can be directly read by the FPGA or with an oscilloscope. An example of the discriminator output is shown in Fig.~\ref{fig:discri} (left) for two input pulses arriving at 20~ns and 30~ns apart, proving the capability of separating signals below the $T_{\rm{TrigDeadTime}}$. The only limitation comes from the discriminator signal width (\ref{fig:discri}, right) which can increase with the input charge from 12 ns up to 24~ns. By identifying the two edges of the signal in the FPGA, we can count the actual number of input signals and retrieve the signal charge (up to 1~pC, i.e. 2 PE for a gain of 3$\times 10^6$) based on the discriminator signal width. It is thus a good complement in the trigger dead time region and provides independent check for the charge acceptance in Section~\ref{sect:shaper}. 

\section{Charge measurements for liquid scintillator applications}\label{sect:charge}

A major difference in the use of CATIROC for liquid scintillator based detectors is that the expected signals are spread in time (up to hundreds ns) compared to the Cherenkov ones. In the previous section we have pointed out a trigger dead time, the  $T_{\rm{TrigDeadTime}}$, affecting consecutive hits arriving within less than $\sim$90~ns. Moreover, we mentioned that the charge is measured from the digitized value of the slow shaper (SSH) amplitude at the peaking time. The slow shaper is an RC filter of second order with a bi-polar signal which may introduce distortions in the charge measurements for pulses arriving close in time.  A dedicated validation is thus needed to ensure the  capability of CATIROC to detect all hits and to retrieve the full charge without any significant bias. In this section, we thus initially prove the charge linearity following the classical approach with input pulses of increasing amplitude. We then inject a train of two pulses performing a scan of their time separation from hundreds of ns to 15~ns. This test mimics the case of two hits arriving close in time and allows to investigate all possible charge distortion introduced by the combination of the SSH shape and the trigger dead time, for different RC values of the SSH. For some of these effects we propose a correction method or we comment on the impact on physics measurements. 

\subsection{Charge linearity}
The linearity of the charge measurement has been tested for all channels for signals ranging between 0.1 and 70 pC. 
Fig.~\ref{fig:charge} shows one example channel, for the HG (red) and the LG line (blue) and for the two SCA capacitors, ping (filled markers) and pong (empty markers). Measurements repeated on a second test board and using an independent setup, provide compatible results in terms of resolution and LSB (Least Significant Bit, given in fC/ADCu).  As reference, the input charge scale is also given in PE units, assuming a PMT gain of $ 3 \times 10^6$ in the charge-to-PE conversion. The results of the linear fit for HG and LG are reported in Table~\ref{tab:lin}, together with a list of the used CATIROC configuration parameters. A measurement of the pedestal is independently performed by measuring the charge distribution with an external trigger (generated by the FPGA): the mean charge is consistent with the intercept (p0) in Table~\ref{tab:lin} and the RMS is about 1.5 ADCu, which provides a signal over noise ratio (SNR) around 40 for a PMT gain of $ 3 \times 10^6$ and the calibration parameters in the table. 
The RMS of the charge distribution has been measured equal to 2 ADCu when operating in HG mode and 1 ADCu in the LG case, which corresponds to 0.015 pC and 0.7 pC, respectively. These charge resolutions correspond to ~3\% for a gain of 3$\times10^6$ (i.e. 0.48 pC) and are within typical PMT-based experiments requirements for which single photo-electron resolutions are expected around 30\%~\cite{SPMT-paper}.  

\begin{figure}[tbh]
\centering
    \includegraphics[width=0.7\textwidth]{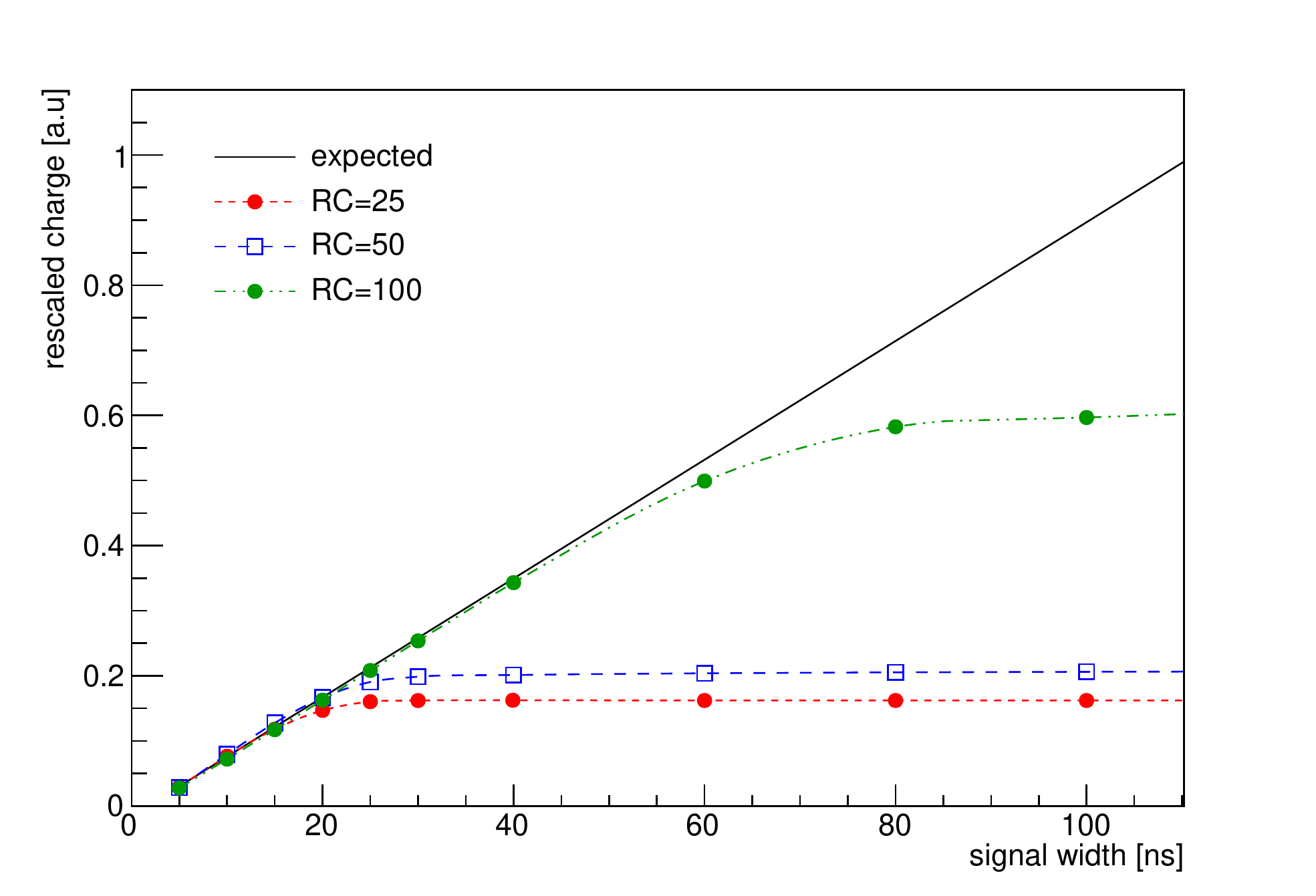}
\caption{Charge linearity as a function of the signal width for different RC constants of the slow shaper. }
\label{fig:width}
\end{figure}

The calibration curve shown in Fig.~\ref{fig:charge} is obtained in the case of RC=50 and a preamplifier gain of 20.  The typical signals of a 3-inch PMT have a width of 10 to 20~ns. The impact of the RC constant  on the integration window can be visualized in Fig.~\ref{fig:width}, where we also tested the charge linearity against the signal width for different RC. For a fixed amplitude of the injected signal, we increase the width of the pulse and we measure a deviation from linearity starting at about 20, 25 and 55~ns, respectively, for the three considered RC constants. For these configurations we checked that no cross-talk signals where observed in other channels than the injected one, over a million of triggered pulses and up to $\sim$60~pC input charges. \\

\vspace{+0.5cm}
\begin{minipage}[t!]{.56\textwidth}
\hfill\includegraphics[width=1.05\textwidth]{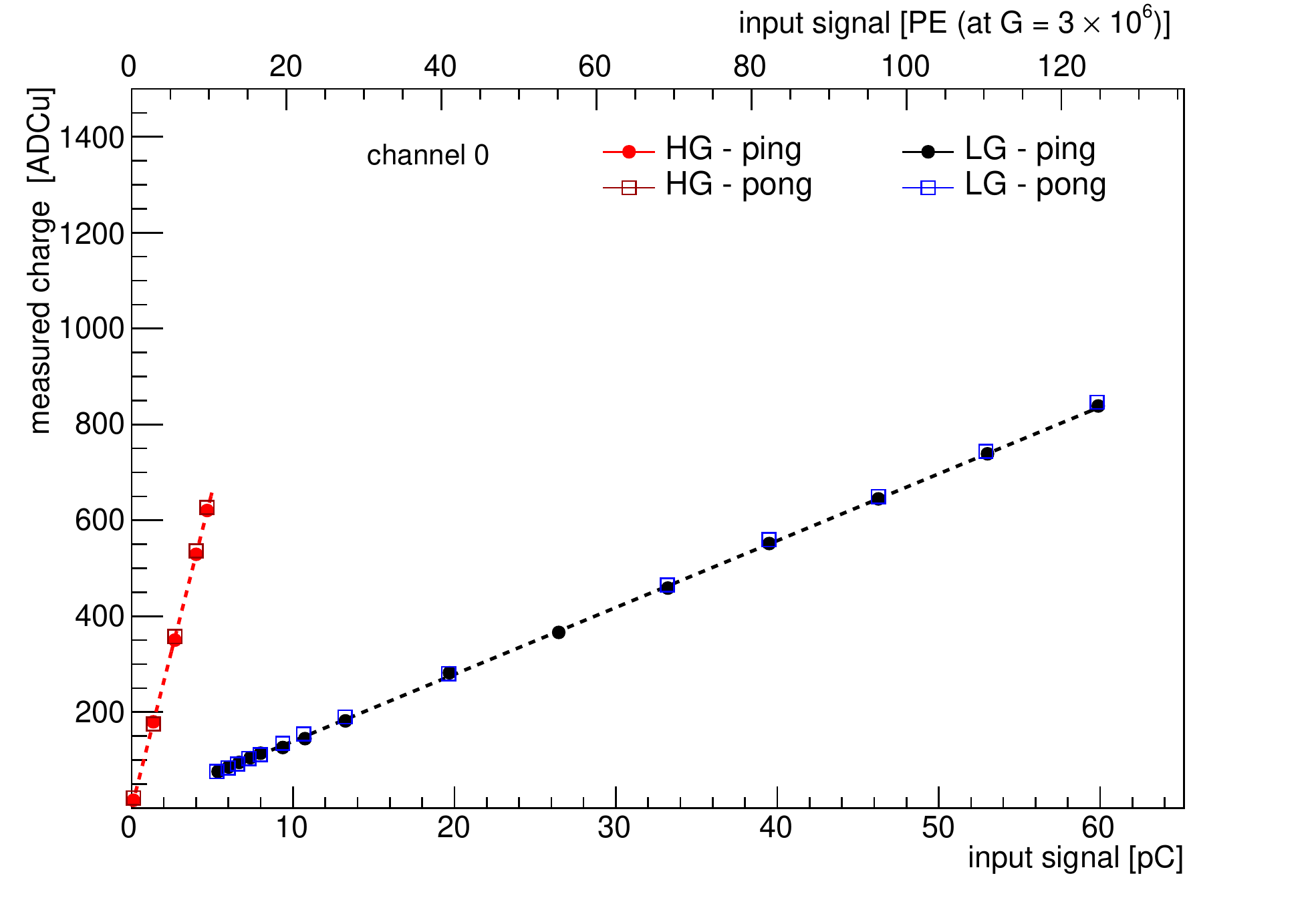}
\captionof{figure}{Linearity of one of the channels for the high-gain in red, and the low-gain in blue. Empty and filled markers denote the two capacitors (ping/pong) in the analog memory. }
\label{fig:charge}
\end{minipage}
\begin{minipage}[t]{.45\textwidth}\centering
\centering
\begin{tabular}{c|c}
\hline
\multicolumn{2}{c}{\footnotesize{Configurations}} \\
\hline\hline
\footnotesize{SSH RC} & \footnotesize{50} \\
\footnotesize{PA gain} & \footnotesize{20} \\ 
\footnotesize{Discr. thr} & \footnotesize{900 DACu}\\
\footnotesize{HG/LG thr }& \footnotesize{720 DACu} \\ 
\footnotesize{SSH Peak time}   & \footnotesize{26 ns} \\
\hline\hline
\footnotesize{$\sigma_{\rm{charge, HG}}$}  & \footnotesize{2 ADCu (0.015 pC)} \\
\footnotesize{$\sigma_{\rm{charge, LG}}$ } & \footnotesize{1 ADCu (0.74 pC)} \\
\hline
\multicolumn{2}{c}{\footnotesize{Calibration fit} } \\
\hline\hline
\footnotesize{intercept (p0)} & \footnotesize{[ADCu]} \\
\hline
\footnotesize{HG} & \footnotesize{$66.9\pm0.3$}  \\ 
\footnotesize{LG} & \footnotesize{$68.6\pm0.1$ } \\
\hline
\footnotesize{LSB (1/p1)} & \footnotesize{[fC/ADCu]} \\
\hline
\footnotesize{HG} & \footnotesize{$7.9\pm0.4$}\\ 
\footnotesize{LG }&  \footnotesize{$73.6\pm 0.1$} \\ 
\hline
\footnotesize{residuals} &  \footnotesize{$<$ 4 ADCu} \\
\hline
\footnotesize{variations} \\ \footnotesize{within channels} & $<$ \footnotesize{5\%} \\
\hline\hline
\end{tabular}
\captionof{table}{Summary of the CATIROC configurations and of the results of the charge linear fit. }
\label{tab:lin}
\vspace{+0.5cm}
\end{minipage}

\begin{figure}
\centering 
    \includegraphics[width=0.7\textwidth]{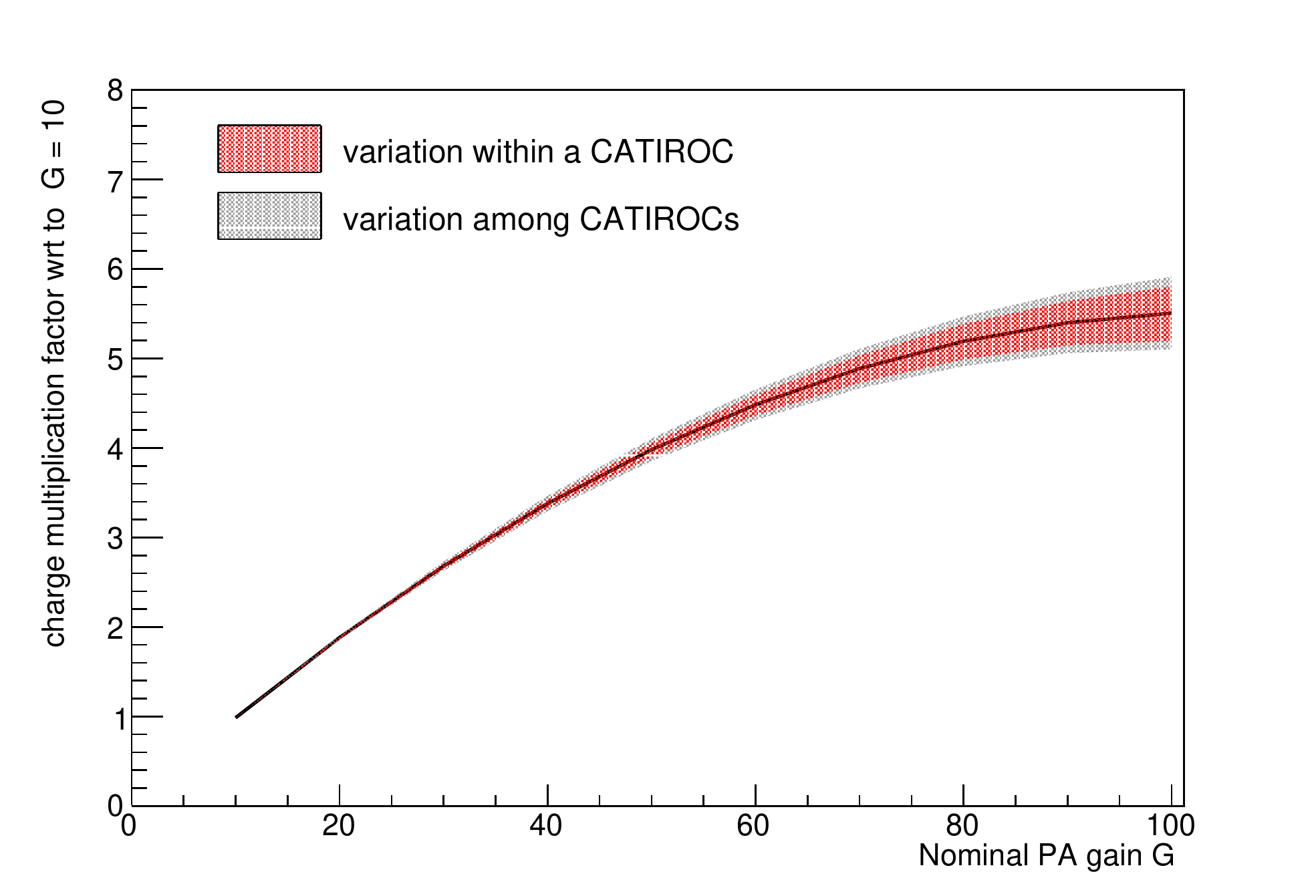}
\caption{Charge multiplication factor (effective gain) vs preamplifier (PA) nominal gain $G$. The average  gain linearity is shown as dark solid line with a gray band indicating the maximum variation between different CATIROCs and a red shaded band indicating the variation within channels of the same CATIROC. }
\label{fig:PAgain}
\end{figure}

For what concerns the preamplifier, its gain directly affects the measured charge as an effective multiplication factor as shown in Fig.~\ref{fig:PAgain}. In the figure the ratio between the measured charge at gain G and the charge measured at the reference case G=10 is plotted against the nominal gain.  Even if the gain can extend more, only the interval up to 60 looks exploitable with small saturation effects. This is sufficient for typical applications where the PMTs can be grouped to have similar gains at the nominal high voltage value.  For completeness, we also indicate the maximum variations observed among three tested CATIROCs (gray band) and within the 16 channels of a same CATIROC (red). In practice, the effect of this variation is negligible as the preamplifier can be fully characterized for each ASIC separately during the production test phase of the chips or of the readout boards.  

\subsection{Slow Shaper effects on charge integration} \label{sect:shaper}
\begin{figure}[t!]
\centering
  \includegraphics[width=0.7\textwidth]{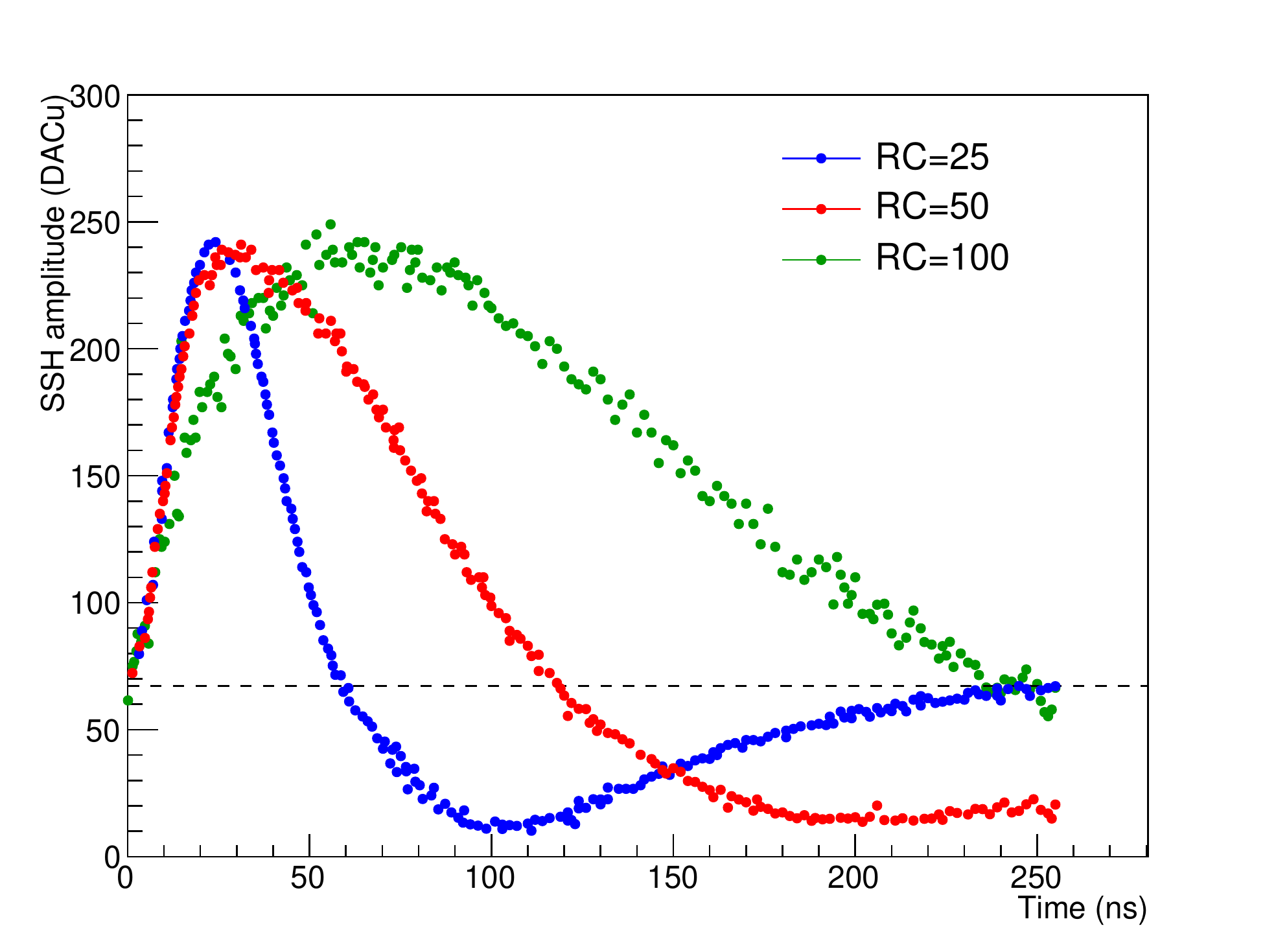}
    \includegraphics[width=0.7\textwidth]{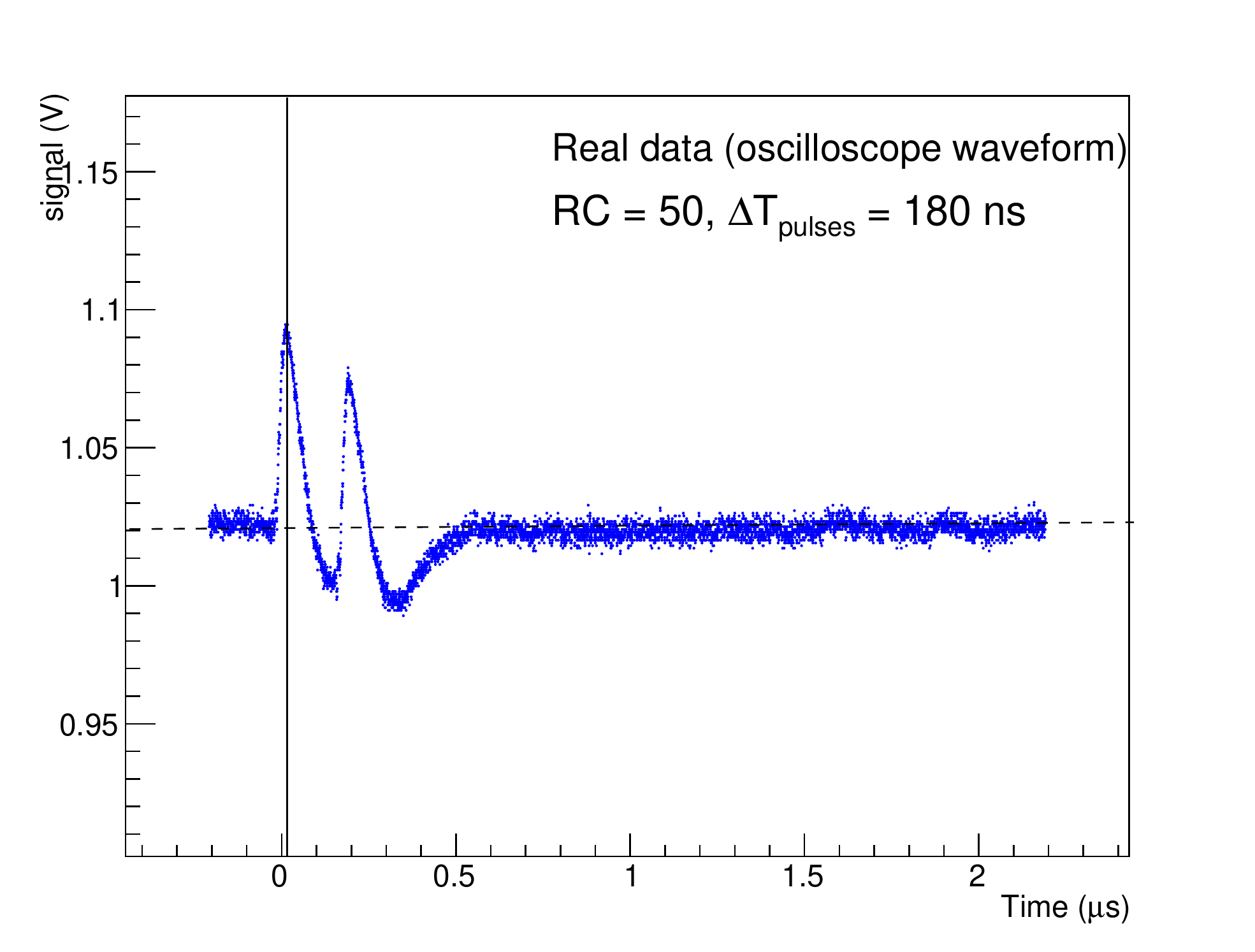}
  \caption{Top: Scan of the SSH vs time (measured from the digitized charge value) for three different RC values zoomed around the peaking time. The undershoot tail starts (ends) around 50 (250~ns), 100 (400 ns) and 250 (1000~ns)  for RC=25, 50 and 100, respectively.  Bottom: case of two signals arriving close in time with the two shapers overlapping. The waveforms are measured with an oscilloscope from one of the probe outputs available for debugging on the evaluation board. The amount of overlap and the possible effects on the measured charge (pile-up, signal loss or signal underestimation) depends on the RC shape and on the $\Delta T$ between the two pulses. }
  \label{fig:SSH}
\end{figure}

As previously mentioned, the charge is measured  from the digitized value of slow shaper (SSH) amplitude at the peaking time (see Fig.~\ref{fig:SSH}, top). In this section we study the possible effects introduced by the slow shaper for three possible configurations of RC (25, 50, 100).   One of these effects is due to the undershoot visible in the figure: a dashed line is drawn to indicate the expected baseline.
We show in Fig.~\ref{fig:SSH} (bottom) the example case of two identical pulses arriving at $\Delta T = 180$~ns: the two SSH overlaps and the amplitude of the second shaper is biased by the undershoot tail of the first one, producing an underestimation of the charge. A second possible bias, which may potentially engender a signal loss, occurs when two signals arrive with a time separation of the order of the  $T_{\rm{TrigDeadTime}}$ or smaller. 

\begin{figure}[h!]
\centering
   \includegraphics[width=0.7\textwidth]{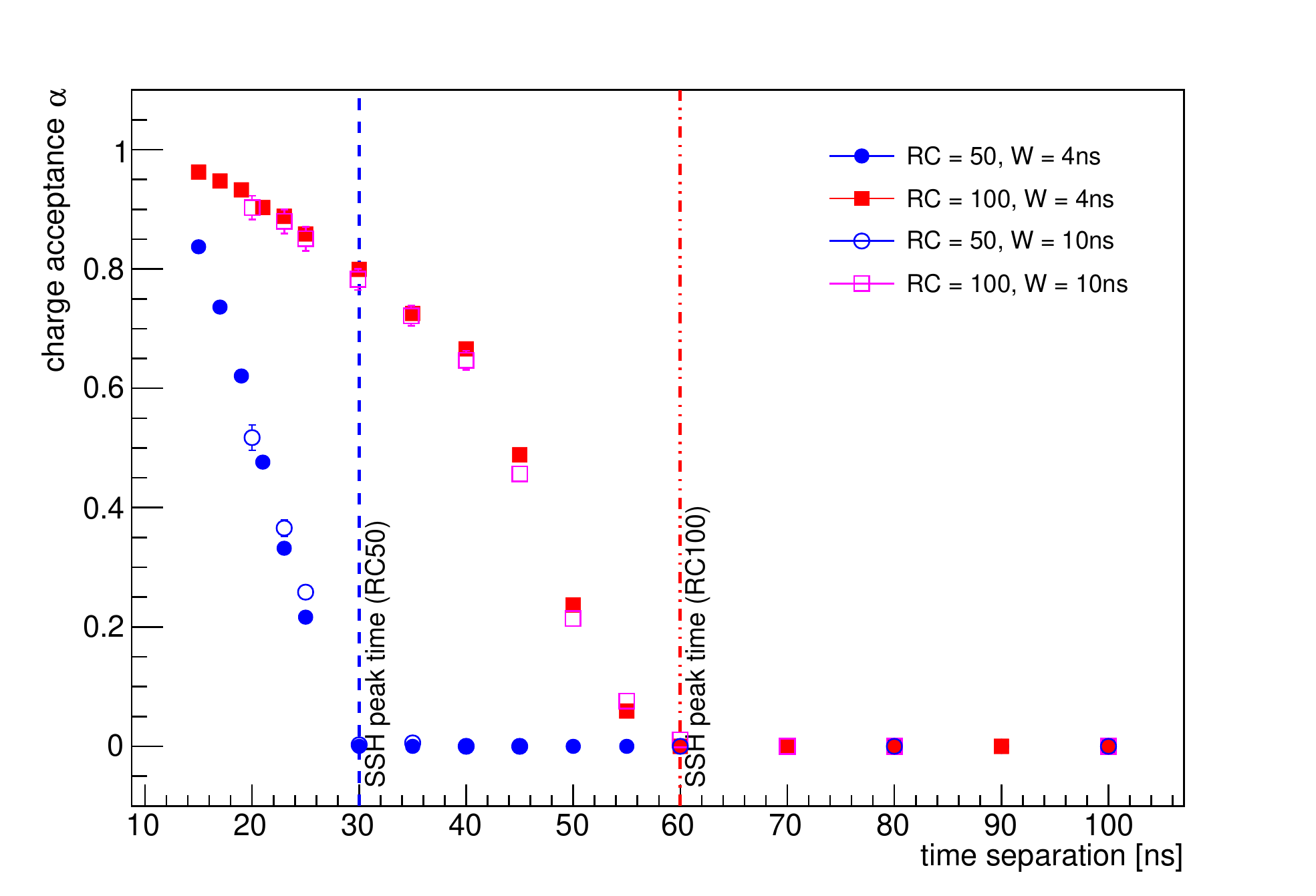}
   \caption{Charge acceptance for RC=50 (circle) and RC=100 (square) as a function of the separation time between pulses. Empty (filled) markers refers to points taken with a signal width of 10~(4)~ns.The dashed lines indicates the SSH peak time for both RC values.  }
   \label{fig:chargeAcceptance}
   \end{figure}
   
\begin{figure}
   \centering
   \includegraphics[width=0.7\textwidth]{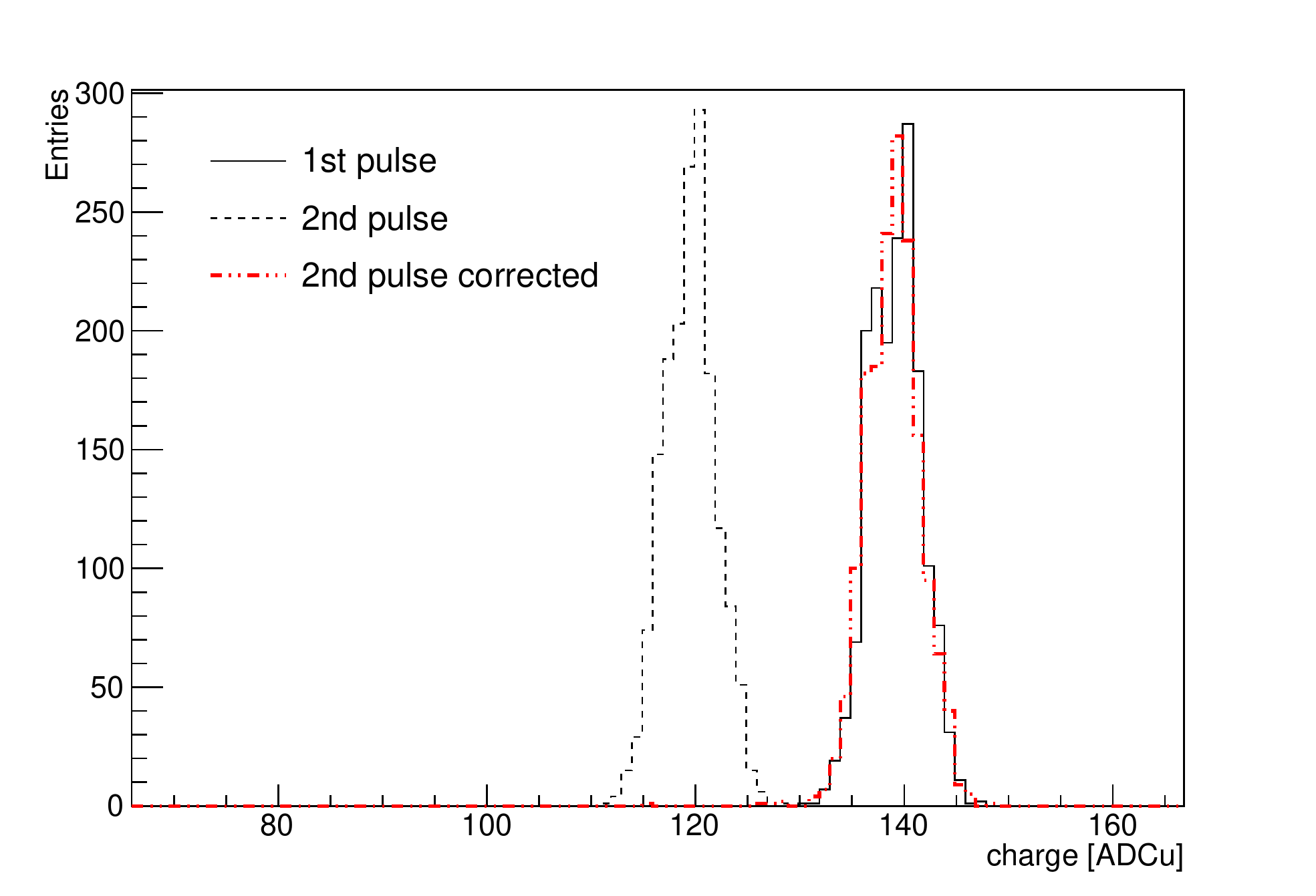}
  \caption{Charge distribution for first hit (solid line) and second hit (thin dotted), for $\Delta T = 100$~ns. The red dashed line shows the charge of the second hit after applying the correction for the bias introduced by the SSH undershoot (see text).  }
  \label{fig:undershoot}
\end{figure}

For the studies presented in this section, we inject a burst of two identical pulses  with a period P~=~1~ms and a time separation between pulses $\Delta T$ varied from 15 to 1000~ns\footnote{The signal generator does not allow for shortest signal thus limiting the pulse separation to twice the signal width}. 
To simplify the description, we indicate with $q^*_1$ and $q^*_2$ the charge measured by CATIROC and $q_1= q_2$ the ``true'' corresponding values. We distinguish two cases :  
\begin{itemize}
\item  $\Delta T \leq  T_{\rm{TrigDeadTime}}$ : only the first hit of each burst is triggered. Its measured charge $q^{*}_1$ may include a fraction of the second charge of the second pulse :  
\begin{eqnarray}
q^{*}_1 &=& q_1 + \alpha(\Delta T,\rm{RC}) \cdot q_2\\ 
q^{*}_2 &=& 0  \nonumber
\end{eqnarray}
with $ 0 \leq \alpha(\Delta T,\rm{RC}) \leq 1$ (hereafter referred to as ``charge acceptance'').  
This fraction $\alpha$ will depend on the time overlap between the two pulses and on the RC shape. It is shown in Fig.~\ref{fig:chargeAcceptance} for RC=50 and RC=100\footnote{The case RC=25 is not considered here since the signal generator would not allow to investigate pulses separations much smaller than the  $T_{\rm{SSHpeak}} \sim 18$~ns.}. The empty markers refers to injected signals with the same width (10~ns) used so far. However, to explore the time separation $<$ 20~ns, we additionally tested the case of signal width of 4~ns.  For signal separations smaller than 15~ns we extrapolate to $f=1$ for $\Delta T=0$ , assuming a function which is constrained by the linearity studies in Section~\ref{sect:charge} (Fig.~\ref{fig:charge}). An important consequence visible in figure is that all the hits arriving with $T_{\rm{SSHpeak}} < \Delta T <  T_{\rm{TrigDeadTime}}$ will not be detected and at the same time their charge does not modify the peak amplitude of the SSH for the first hit (i.e. $\alpha = 0$). 

\item $\Delta T >  T_{\rm{TrigDeadTime}}$ (\emph{undershoot bias}) : The two hits are triggered independently but, as illustrated in Fig.~\ref{fig:SSH} (bottom), the undershoot acts as an offset for the second SSH. This induce an  underestimation of the measured charge ($q^*_2 < q_2)$: 
 \begin{eqnarray}
q^{*}_1 &=& q_1 \nonumber \\
q^{*}_2 &=& q_2  -  \kappa (\Delta T,\rm{RC})*q_1  
\end{eqnarray}
where $\kappa (\Delta T,\rm{RC}) < 0$  models the SSH signal (in ADCu)  normalized to a peak amplitude equal to one.  Since the SSH normalized shape is preserved (for a given channel and RC), for each hit we can retrieve the ``true'' charges correcting for the modeled $\kappa$. This correction is working perfectly as illustrated on Fig.~\ref{fig:undershoot}. 
\end{itemize}

\section{Time resolution}\label{sect:time}

\begin{figure}[t!]
\centering
  \includegraphics[width=0.7\textwidth]{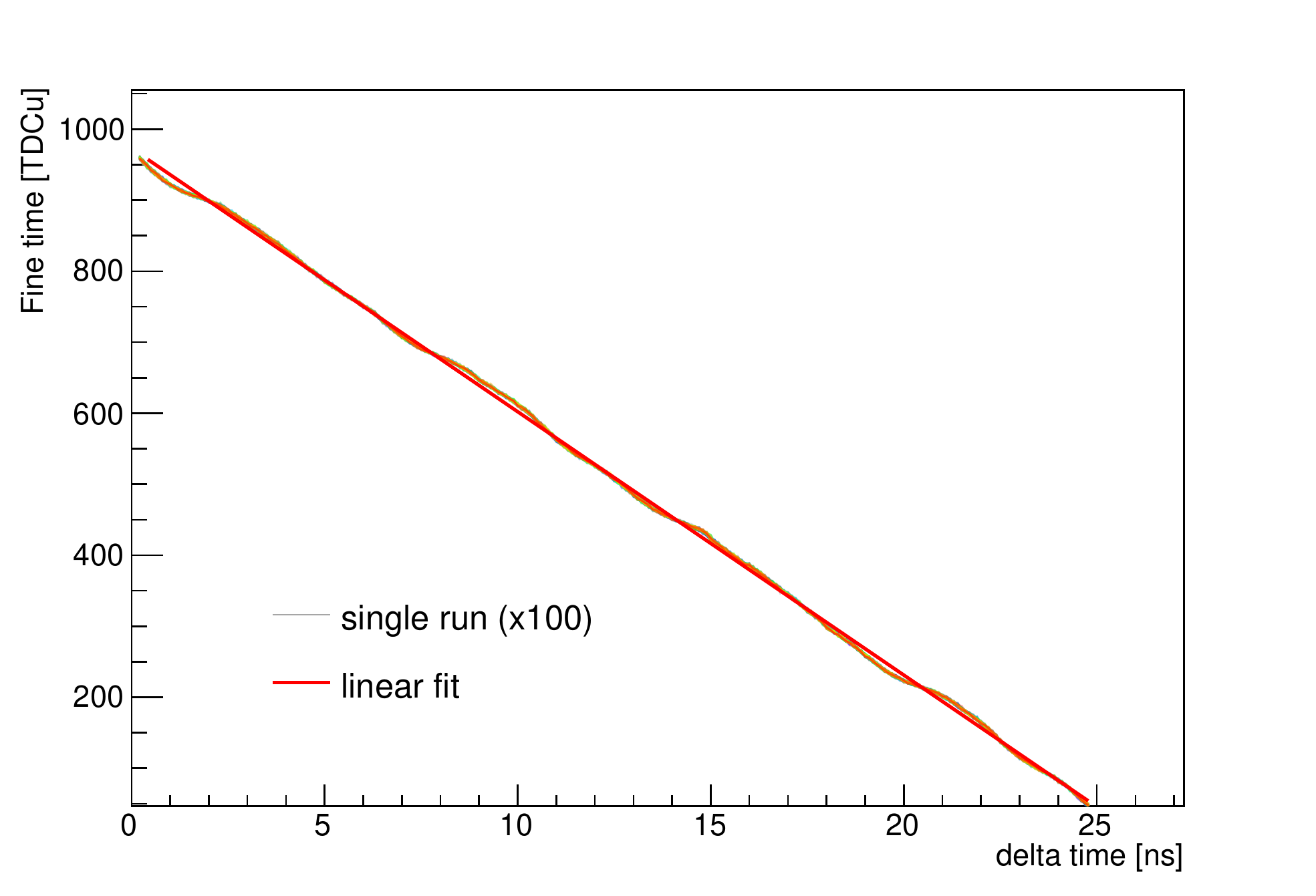}
  \includegraphics[width=0.7\textwidth]{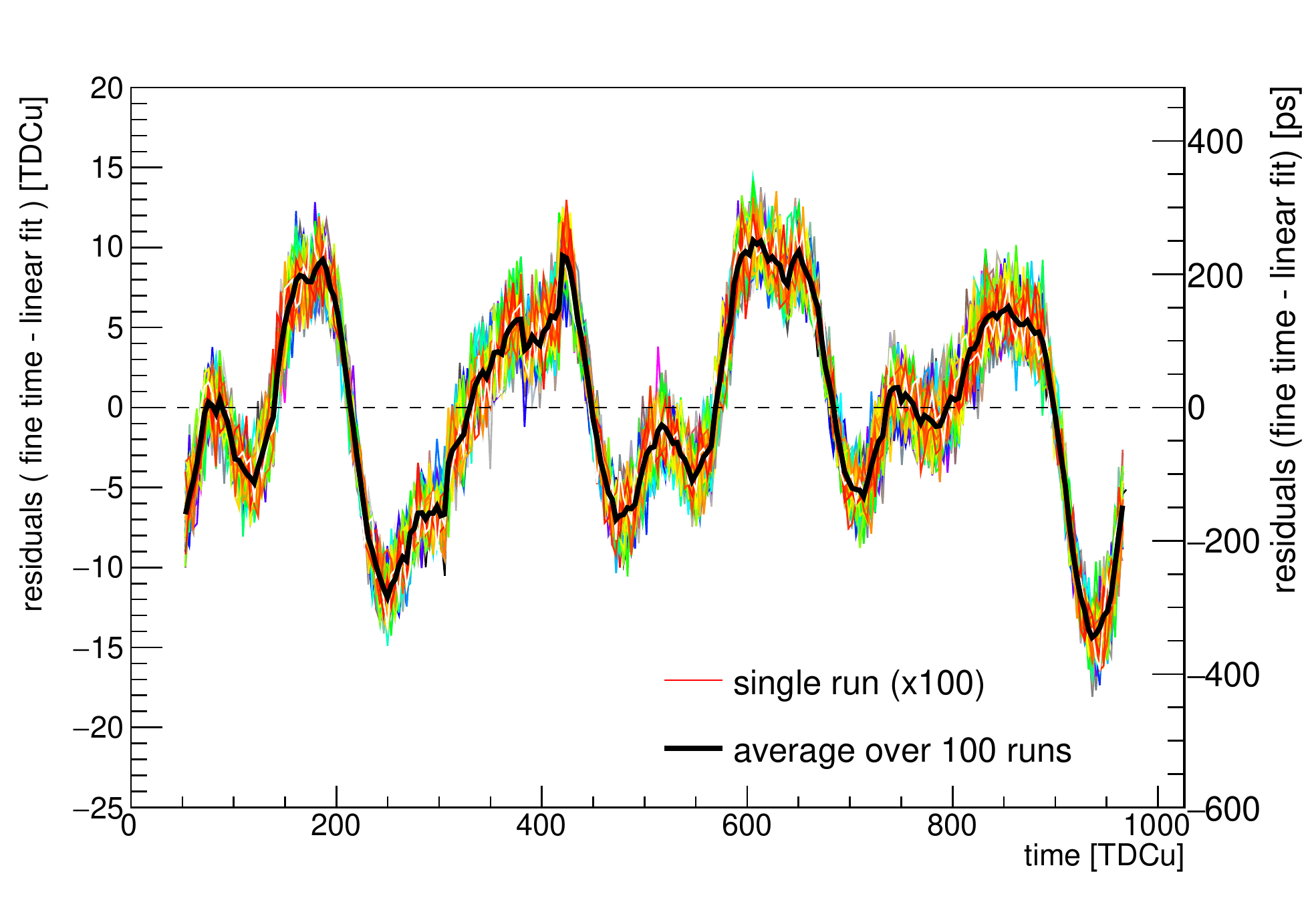}
  \caption{Top: TAC response (after digitisation) obtained from a scan in time with step size of 100 ps. The measurement is repeated for 100 runs. Bottom: Residuals with respect to the fit of each scan curve (thin colored lines) and for the average value (black solid).}
  \label{fig:TDCscan}
\end{figure}

As described in Section~\ref{sect:catiroc}, the time measurement is the combination of a 26-bit coarse time and a fine time measured with a TAC and digitized by a 10-bit Wilkinson ADC. A scan of the TAC ramp is done spanning the clock cycle (25~ns) in steps of 0.1~ns as shown in Fig.~\ref{fig:TDCscan} (top). 
An example of the TDC measured values (thin gray line) and the  linear fit (thick red line) is  shown in the top panel.  The parameters of the linear fit to the average over hundred independent runs, gives $973.2 \pm 0.1$ TDCu and $-37.10 \pm 0.01$ TDCu/ns for the intercept and the slope, respectively. For each measurement,  the residuals from the fit function distribute with an  RMS (over the full time window) of about 0.15~ns. Moreover, these residuals (thin colored lines in Fig.~\ref{fig:TDCscan}, bottom) exhibit a modulation due to the clock coupling in the substrate. The average curve (solid thick line) ranges between $^{+10}_{-15}$ TDCu, which corresponds to  $^{-0.27}_{+0.41}$~ns, with variations of $\pm 2$ TDCu ($\sim$~54 ps) between different runs. 
As emphasized by the figure, the observed modulation is well reproducible between different runs and may be corrected offline, which is an interesting possibility for specific applications aiming at a time resolution below 100~ps. 
Such a study is out of the scope of this paper, because they are much smaller than the typical PMT  time transit spread ($\sigma$) which is typically of 1.5-2~ns~\cite{SPMT-paper, hzc}.
In the validation board used for this paper, the voltage levels of the TAC ramp are not well adapted to the ADC input ones  and the digital output spans a range smaller than the available bits (970 ADCu instead of 1024 ADCu). 
This reduces the LSB from the nominal value of 24 ps/TDCu to 27 ps/TDCu.  Though it  has negligible effects for the results shown here and for systems dominated by PMT resolutions,  this non-adaptation could be solved by adding an external resistor on the board.
\begin{figure}[htb!]
\centering
\includegraphics[width=0.7\textwidth]{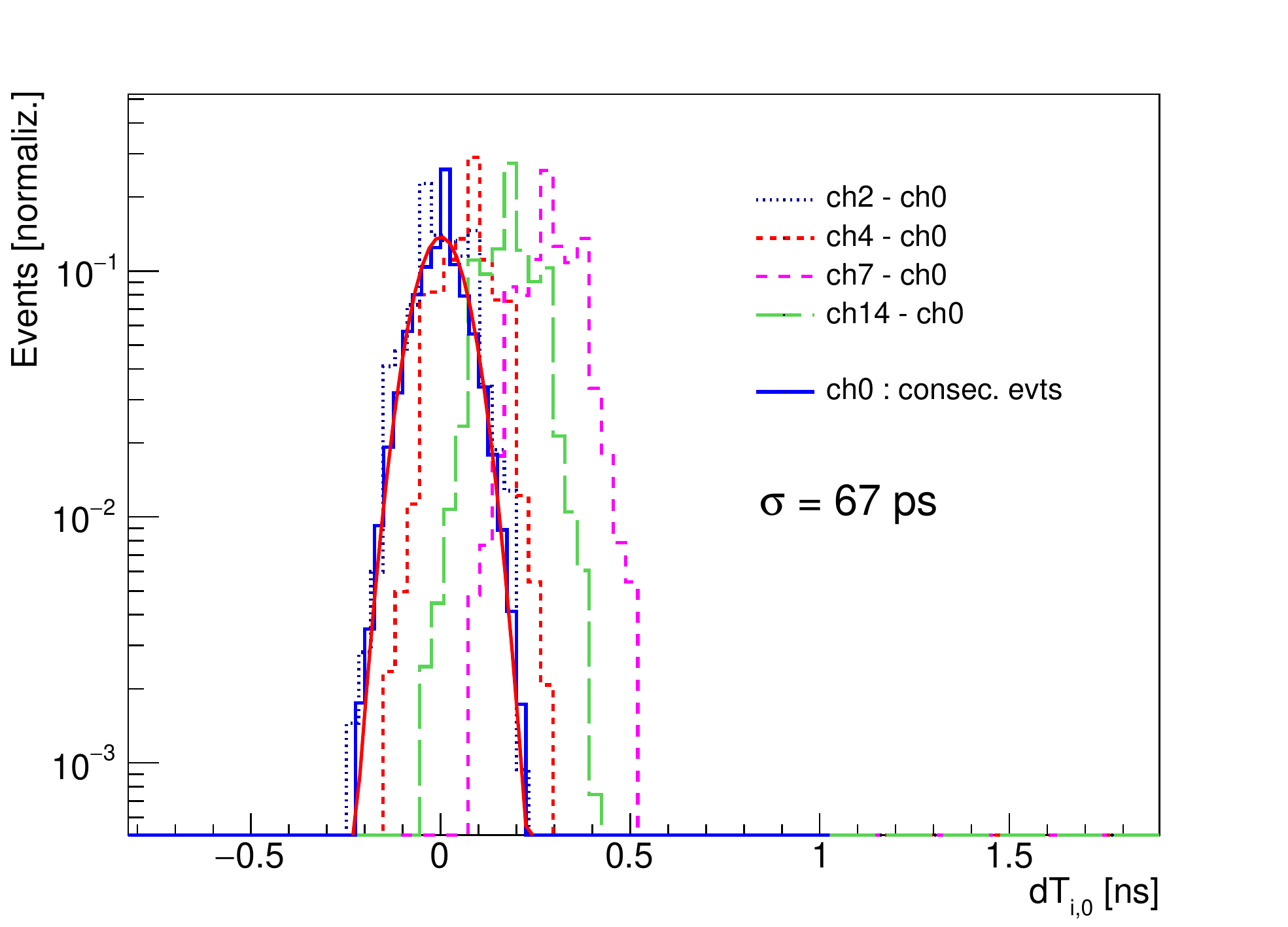}
\caption{Example distributions of $dT_{i,j}$ between the $i$-th channel and channel  $j = 0$ (thin dashed lines). The thick blue histogram is the case of $dT_{\rm{00}}$ (subtracted of the signal generator period) obtained from the time difference between consecutive signals injected in the same channel ($i = j = 0$). }
\label{fig:time2}
\end{figure}

A check of the time resolution ($ \sigma_{\rm{T}}$) of the system is also obtained by measuring the time difference, $dT$, between two consecutive signals injected in the same $i$-th channel and the time difference $dT_{ij}$ measured for the same  signal injected in the channels $i$ and $j$ using a passive signal splitter (FANOUT).  
In the first approach, the $dT$ distribution is expected to be centered at the signal generator period, with a spread that is related to the time resolution convoluted with the jitter of the signal generator. An example is shown in Fig.~\ref{fig:time2} where the thick blue distribution represent the case of channel 0. The signal generator period is subtracted for plotting reasons. The distribution is fitted with a gaussian function giving a  $\sigma$ of 67 ps. This results is obtained similarly for all channels and is mostly affected by the modulation of the TAC value mentioned above.  
The second approach is introduced to get rid of the systematics due to the signal generator, even if it will be then sensitive to possible systematics between different channels. The distributions of $dT_{ij}$ in Fig.~\ref{fig:time2} are then expected to be centered at zero with a measured spread ($\sigma_{ij}$) given by the time resolutions $\sigma_{i}$ and $\sigma_{j}$ of the two channels by: 
$$\sigma_{ij} = \sqrt{\sigma^{2}_{i} + \sigma^{2}_{j}} \approx \sqrt{2} \sigma_{\rm{T}}$$ 
under the assumption that $\sigma_{i} \approx \sigma_{j}= \sigma_{\rm{T}}$. We conservatively take the spread $\sigma_{ij}$ as time resolution per channel.  
The distributions $dT_{i,0}$  ($i \neq 0$) are not perfectly centered at 0 which can be explained as an effect of the FANOUT (tested by mixing the channels in inputs of the FANOUT) and a dispersion of the TAC ramps of different channels.  
With both approaches, the value of $\sigma_{ij}$ is within 150 ps, compatible with the results from the TAC ramp residuals.

\begin{figure}
\centering
\includegraphics[width=0.7\textwidth]{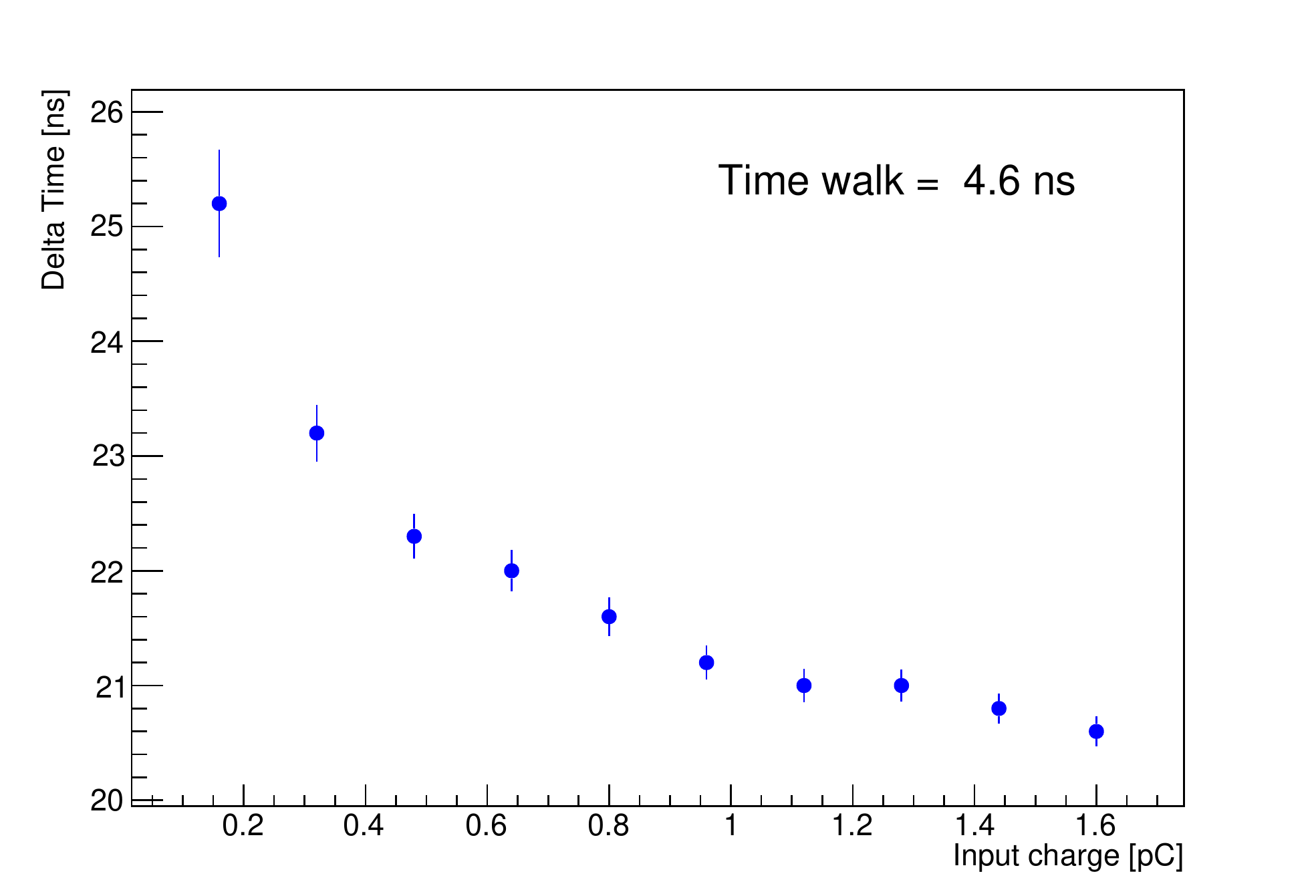}
\caption{Difference between the trigger time and the pulse generator signal as a function of the input charge.  A time walk of 4.6~ns is measured, for a trigger threshold of 160 fC (about 0.3 PE at $3 \times 10^6$ and preamplifier gain of 20). }
\label{fig:timewalk}
\end{figure}

The accuracy of the time measurement is limited by the ``time walk'' of the fast shaper, which is basically the time at which the fast shaper signal exceeds the trigger threshold. The time walk depends mostly on the signal amplitude. In this study the time walk has been measured injecting a charge with a pulse generator and comparing the trigger time in CATIROC and the signal time. 
Fig.~\ref{fig:timewalk} shows the difference between these two times as a function of the injected charge for a trigger threshold of 0.3 PE. In this configuration, the time walk of the CATIROC circuit has been measured to be around 5 ns. It should be noted that the time walk depends on both the PA gain value and the DAC threshold value and that a correction of the time walk can be applied based on the input rise time and the measured charge.  


\section{The single photo-electron spectrum with PMTs}\label{sect:spmt}

\begin{figure}[t!]
\centering
  \includegraphics[width=0.7\textwidth]{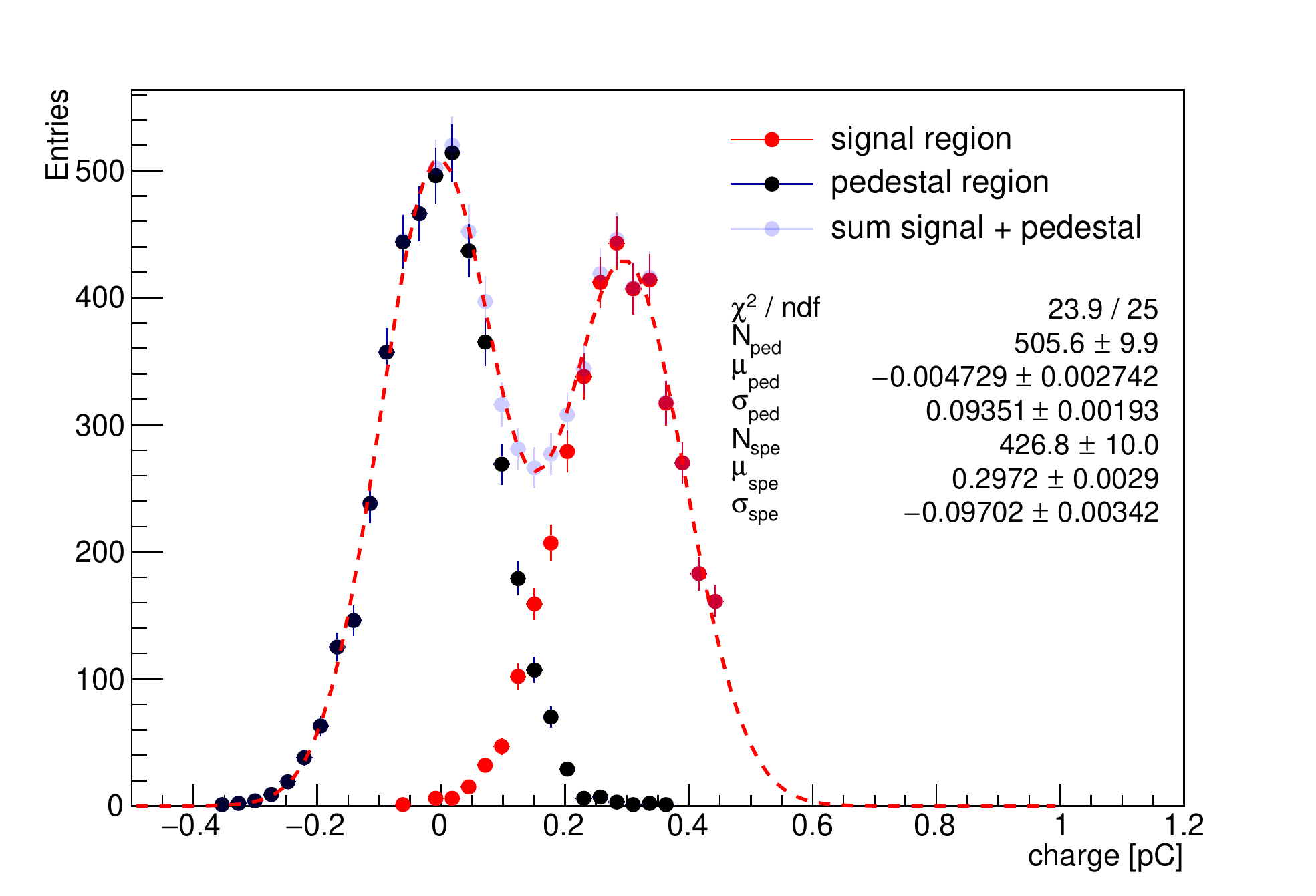}
  \includegraphics[width=0.7\textwidth]{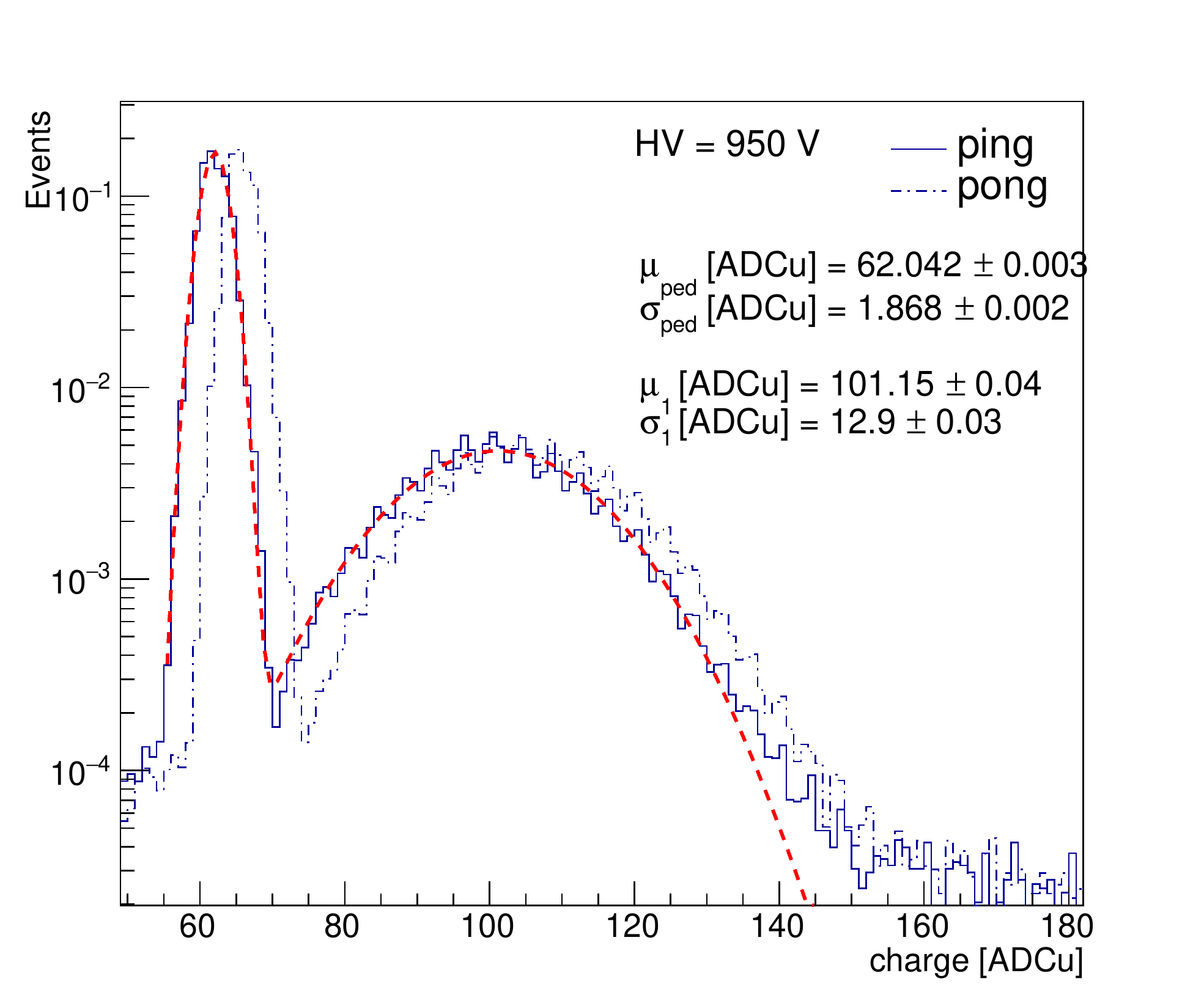}
\caption{Top: Example of the PMT spectrum measured with an oscilloscope: in black and red the charge integrated in the pre-trigger (baseline) and in the signal regions (PE), respectively; in light blue the sum of the two distribution. The fit with the function~(\ref{eq:fitfnct}) is shown as dashed red line. Bottom: example of the spectrum of a 3-inch PMT measured with CATIROC. Data acquired with the ping and pong capacitors can be reconstructed separately guaranteeing that the resolutions are not degradated.} 
\label{fig:PMT2}
\end{figure}

In this section we demonstrate the CATIROC performance when employed to read the signals from a 3-inch PMT. The aim of this section is not to provide a complete study of the PMT response but rather to validate the charge spectrum measured by CATIROC and point-out possible effects introduced by the circuit. To validate the ASIC response we firstly measure the single photo-electron (SPE) spectrum with an oscilloscope and then we compared the obtained PE position and resolution with the one measured by CATIROC. 

For this purpose, we set up a test-bench consisting of a light-tight box (50$\times$40$\times$25~cm$^3$) containing up to two 3-inch PMTs (model HZC XP72B22~\cite{hzc}) powered with a negative high voltage (HV).
A simple approach to measure the SPE spectrum is to look at the dark noise, which is mostly due to thermo-ionic emission from the photocathode, leakage current between the cathode and the dynodes and ionization from residual gases. 
To ensure that a steady dark current is reached, data are always acquired at least 3 hours after any manipulation requiring the PMT exposition to light. 

The PMT spectrum measured with an oscilloscope is shown in Fig.~\ref{fig:PMT2} (top),  for the case of a PMT with a  high voltage of -950~V. The first peak (black markers) is obtained integrating the trace in a time window of 50 ns before the trigger and thus provides a measurement of the pedestal. The second peak (red) is the PMT signal integrated in a time window of 50 ns after the trigger.   
The total PMT spectral response, can be described by a simplified function~\cite{bellamy}: 

\begin{equation}
{\rm f}(i)= \frac{N_{\rm Ped}}{\sqrt{2\pi}\sigma_{\rm Ped}} \, \exp\left( -\frac{(S_i-\mu_{\rm Ped})^2}{2\sigma^2_{\rm Ped}}\right) \, + \sum_{j>0}{\frac{N_j}{\sqrt{2\pi}\sigma_j} \exp\left(-\frac{(S_i - \mu_j - \mu_{\rm Ped})^2}{2\sigma^2_j}\right)}\
\label{eq:fitfnct}
\end{equation}

The first term is the gaussian fit of the pedestal, while the second term describes the PE spectrum in its general form, summing over $j>0$ PEs. In the case considered here, the sum reduces to $j=1$ (SPE).  In the equation, $S_i$ is the value of the $i$-th ADC bin, $N_{\rm{Ped}}$, $\mu_{\rm{Ped}}$, $\sigma_{\rm{Ped}}$ are the parameters of the gaussian fit to the pedestal distribution; $N_j$, $\mu_j$ and $\sigma_j$  gives the amplitude, position and spread of the $j$-th photo-electrons. Hereafter, we denote the SPE mean position as $\mu_{\rm{SPE}} = \mu_{\rm{1}}  -  \mu_{\rm{Ped}}$. 
The results of the fit of equation~\ref{eq:fitfnct} to the data are reported in figure, with an SPE position of 0.30~pC (which corresponds to a gain of $1.8\times10^{6}$) and a sigma of 31\%. 

The charge spectrum measurement is repeated using the same PMT and the CATIROC evaluation board\footnote{The slow control parameters are configured as for Fig.~\ref{fig:charge}, with the exception of the trigger threshold that is here set to 950 DACu to observe also the pedestal peak.}.  
\begin{figure}[t!]
\centering
\includegraphics[width=0.7\textwidth]{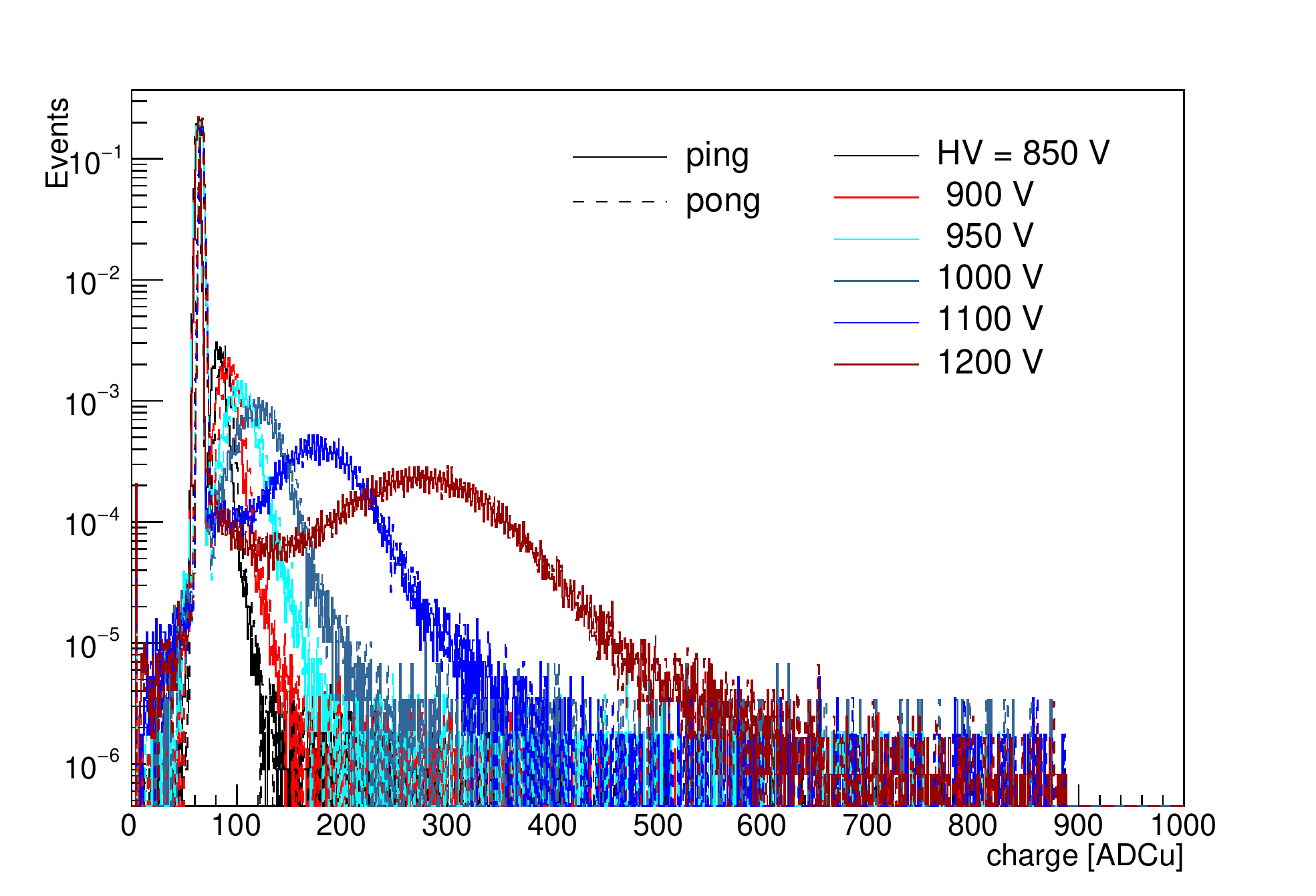}
\includegraphics[width=0.7\textwidth]{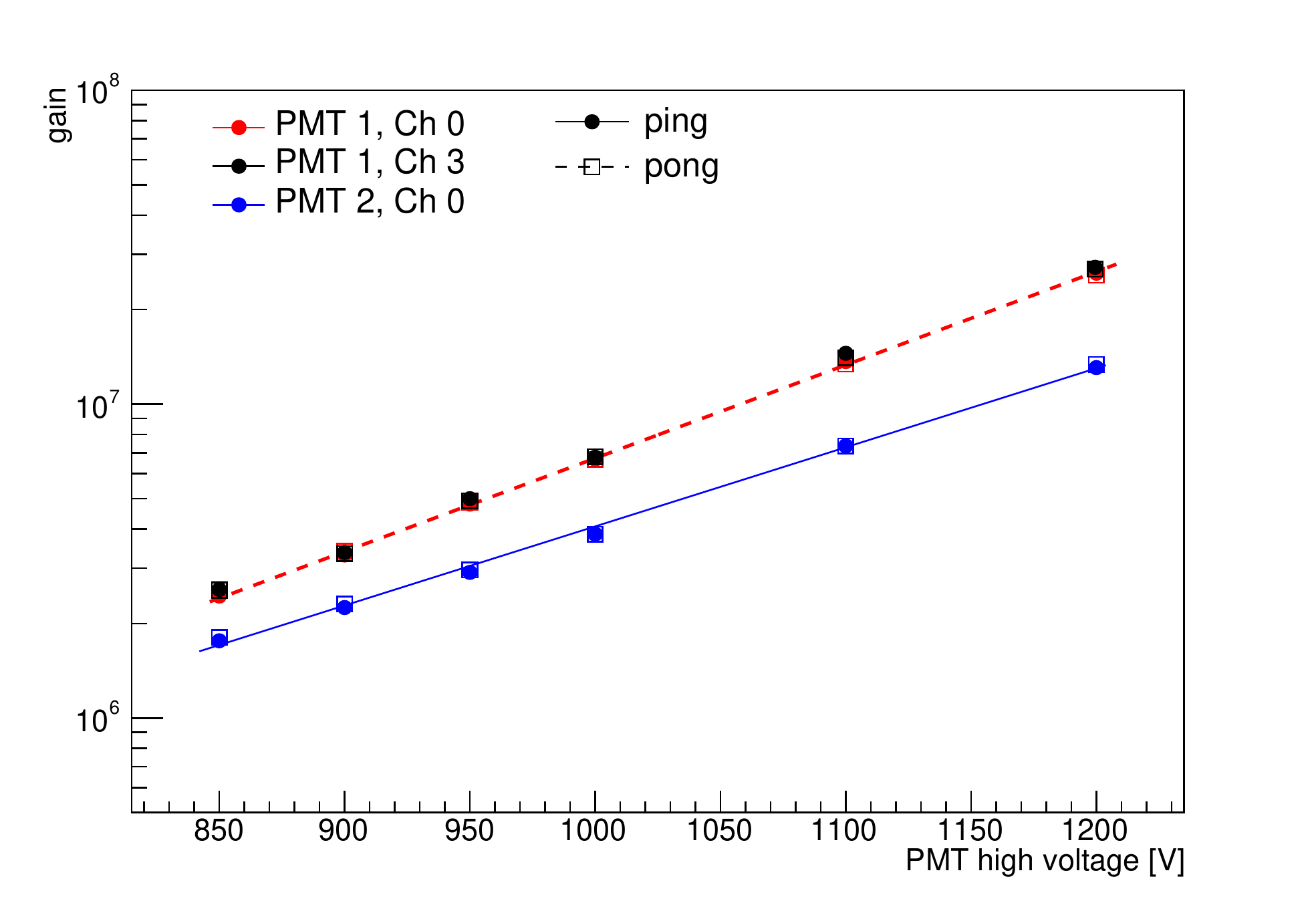}
\caption{Top: charge spectrum measured for different HV.  Bottom: Gain measured with CATIROC for two PMTs. The conversion from ADC to charge is done using the linear fit in Fig.~\ref{fig:charge}. For a given PMT, the measured gain is independent of the selected channel (red vs black) and of the use of ping or pong capacitors (empty and filled marker). For comparison, a second PMT (in blue) is tested using the same channels.}
\label{fig:PMT3}
\end{figure}

The results are shown in Fig.~\ref{fig:PMT2} (bottom) with the fit function (eq.~\ref{eq:fitfnct}, with $j<=1$) drawn as a red dashed line.  The SPE position, calculated after pedestal subtraction, is 39 ADCu, which corresponds to 0.3 pC when calibrated with the LSB in Table~\ref{tab:lin}. The relative SPE resolution is 33\%, well compatible with the result from the oscilloscope. As clearly visible in the figure, the spectral distributions are slightly different for ping and pong data. This difference, mostly due to the offset in the pedestal, is stable against the HV and is corrected once the proper calibration for ping and pong is applied (see Fig.~\ref{fig:PMT3}). 
 \begin{figure}
\centering
\includegraphics[width=0.7\textwidth]{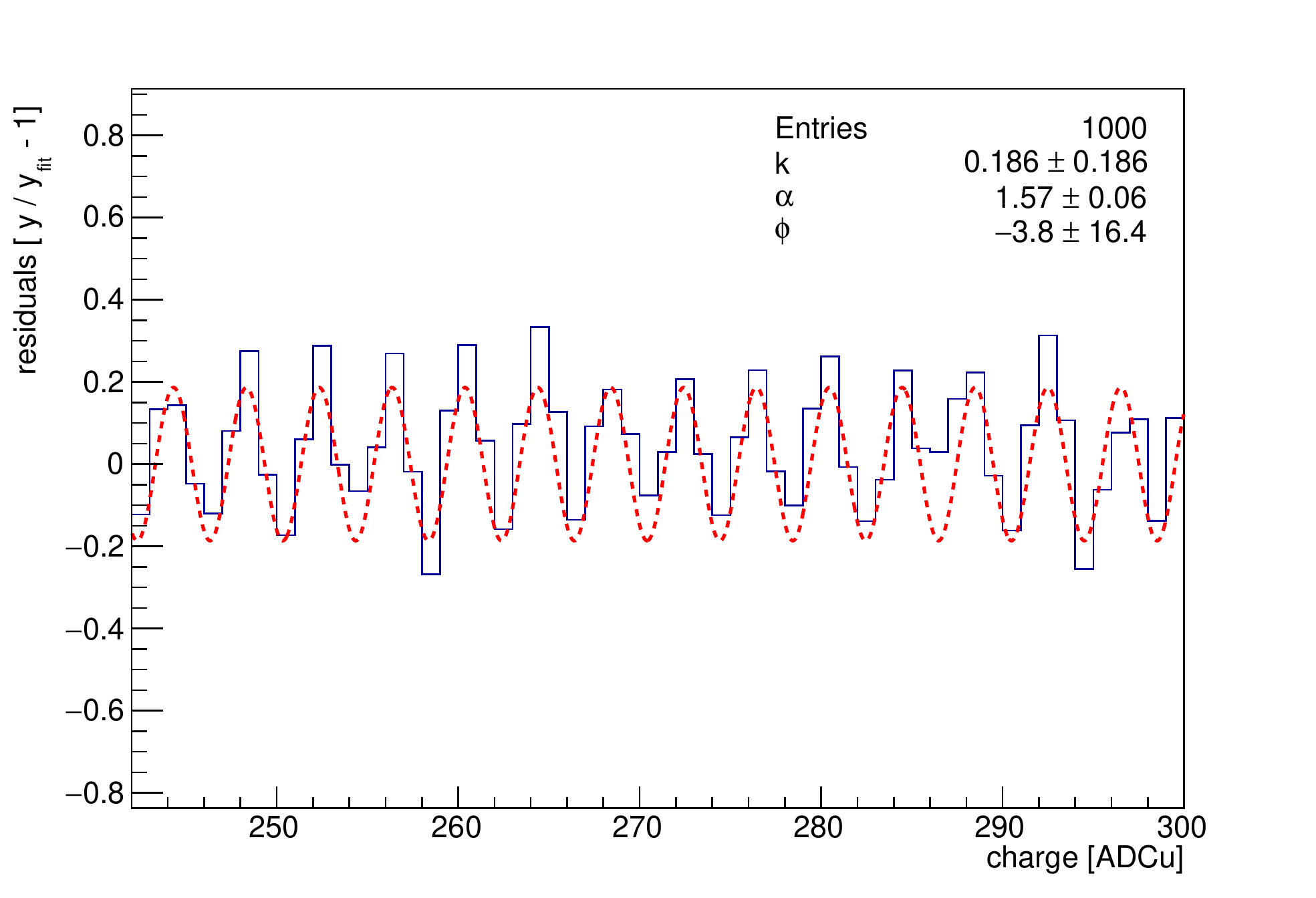}
\includegraphics[width=0.7\textwidth]{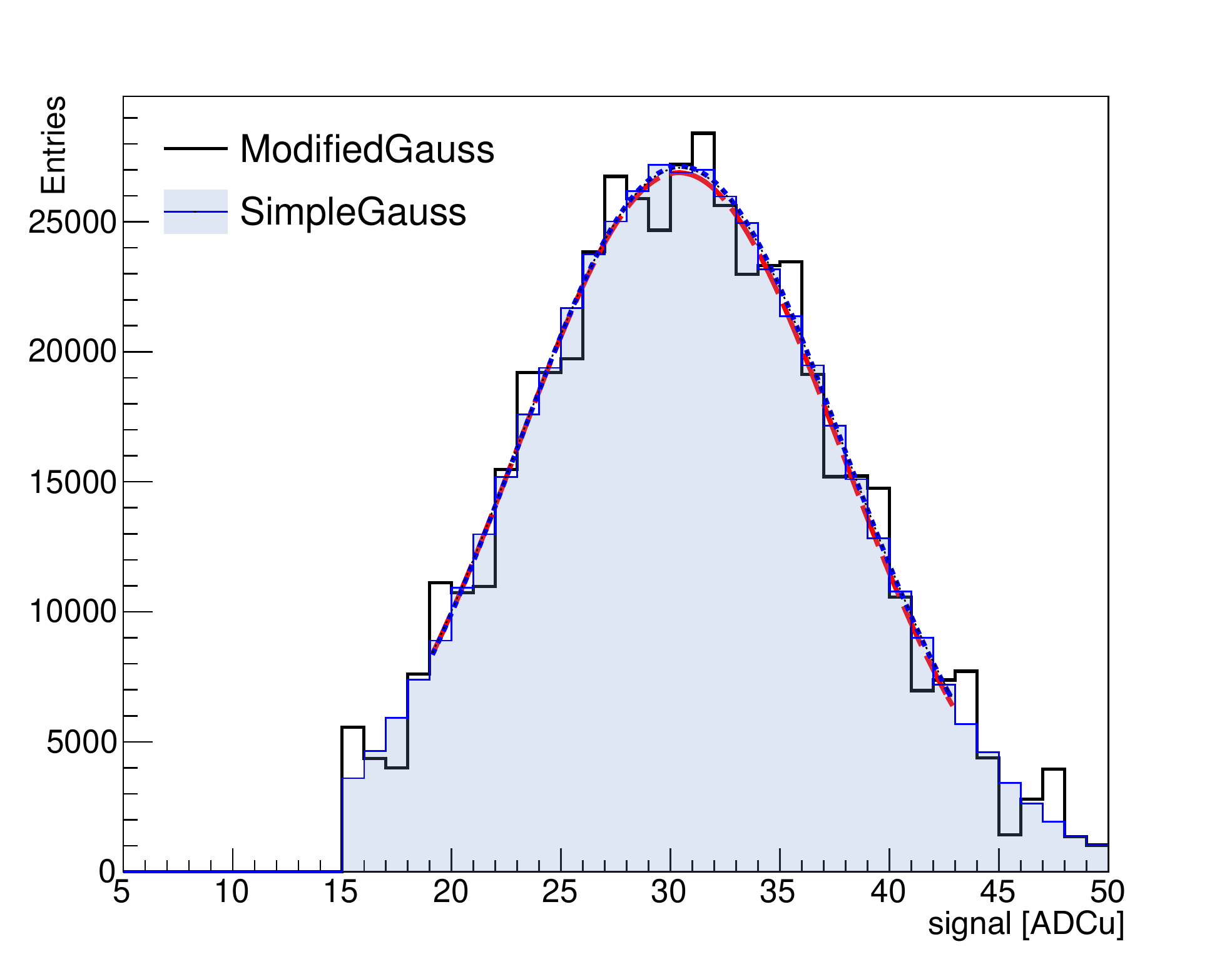}
\caption{Top: Plot emphasizing the wiggles observed in the SPE spectrum and the fit function (red dashed line) with a sinusoid function. Bottom: a toy Monte-Carlo test applying the wiggles modulations on a gaussian SPE distribution, assuming pedestal subtracted. In shaded blue the original distribution and in black after the modulation. The impact on the reconstructed mean SPE position is negligible (compare the two fits as red and blue dashed lines). }
\label{fig:wiggles}
\end{figure}

The SPE spectrum measured for different HV values is shown in Fig.~\ref{fig:PMT3} (top) for one PMT. The SPE mean position ($\mu_{\rm{SPE}}$) and spread ($\sigma_{\rm{SPE}}$) rise with HV as a consequence of the increasing PMT gain. At large HV, the SPE spectrum is better fitted including an exponential tail but we leave this study to a dedicated paper on the PMT performance.

Fig.~\ref{fig:PMT3} (bottom) shows the gain of two PMTs as a function of the applied HV. The measured gain for ping and pong perfectly overlaps after calibration. Moreover, we confirm that the measured gain is the same in different channels using the respective calibrations (red vs black).  
Moreover,  we checked that the pedestal is suppressed for increasing  discriminator threshold without affecting the SPE distribution and we find out that with a threshold of 850 DACu we can cut out all signals with a charge smaller than 70 ADCu.

Finally, we want to discuss briefly  the effect of a clock-coupling which is visible as ``wiggles'' in the measured charge spectrum (Fig.~\ref{fig:PMT2}, bottom). This effect is well reproducible and is emphasized in Fig.~\ref{fig:wiggles} (top) when comparing the bin content in each bin and the fit function in equation~\ref{eq:fitfnct} for the case $j=1$. Its trend can be modeled with a function 
\begin{equation}\label{eq:wiggles}
\frac{N_i}{N_{\rm{fit},i}} = k \times \sin (\alpha x + \phi)
\end{equation}
where $N$ is the measured value and $N_{\rm{fit}}$ the one expected from the fit, and $k$, $\alpha$ and $\phi$ are the amplitude, frequency and phase of the modulation term.  
To evaluate the impact of the wiggles in the determination of the PE position, a toy Monte Carlo test was performed (see Fig.~\ref{fig:wiggles}, bottom). A Gaussian distribution with mean ($\mu_{\rm{true}}$) and standard deviation ($\sigma_{\rm{true}}$) is simulated (blue shaded area). The modulation effect is then applied, using eq.~\ref{eq:wiggles}. The distorted distribution (solid black) is then fit with a Gaussian function and the derived $\mu_{\rm{PE}}$ and $\sigma_{\rm{PE}}$ compared to the fit parameters from the gaussian fit of the true distribution.  A discrepancy smaller than 0.01 ADCu and 0.3\% for $\mu$ and $\sigma$, respectively is obtained and is mostly independent of the $\mu_{\rm{true}}$ and $\sigma_{\rm{true}}$, in the range of the typical values for PMTs.

\section{Summary and discussions}\label{sect:discussions}

The CATIROC ASIC, developed for PMT-based applications, has been tested for its future application in Cherenkov and liquid scintillator experiments.  The main features of CATIROC $-$ the auto-trigger mode, the compressed digital output and the flexibility offered by the slow-control parameters $-$ are of high interests for next-generation experiments which are all facing systems with a large number of channels. 
In such cases, a full readout and storage of the signal waveform (e.g., Flash ADC trace) is impracticable. The small amount of data to be transferred from the CATIROC and the number of input channels in a single chip allows to develop a simplified electronics. An example application of CATIROC is the readout of the 3-inch PMTs by the JUNO liquid-scintillator neutrino experiment, under construction in China~\cite{JUNO-physics}. 

The present paper gives a complete description of the CATIROC features and provides also a first proof of its relevant use in a liquid scintillator experiment. 
Among the major features of CATIROC described here, particular attention have been be payed to the dead times induced by the handling of the triggers ($T_{\rm{TrigDeadTime}}$ in sections ~\ref{sect:deadtime} and ~\ref{sect:shaper}) and by the digitization of the charge ($T_{\rm{DeadTime}} \sim 9\, \rm{\mu s}$ with the mitigation offered by the SCA system).  
For those input pulses not producing an independent charge measurement because of $T_{\rm{TrigDeadTime}}$, we proved  that their charge can be recovered as it summed up to the previous hit according to the charge acceptance in Section~\ref{sect:charge}. Moreover we mentioned the possibility of using the analog signal from the discriminator to count independent triggers, even when arriving within the $T_{\rm{TrigDeadTime}}$. 
The dead times and the effects of the slow-shaper are the most critical points for the application of CATIROC to liquid scintillator detectors. This is doable for PMTs operation in photon-counting mode for most of the physics goals.  
For normal triggered pulses, we demonstrate the charge linearity for each of the input channels for both HG and LG circuits, which ensures a wide operation range, up to 70 pC, without signal saturation.
The achieved charge resolutions are about 15 fC in HG and 73 fC in LG mode. The trigger threshold can be set sufficiently low to measure charges down to 0.16 pC ($\sim$~0.3 PE for a PMT gain of $ 3 \times 10^{6}$) which is the typical choice for PMT-based experiments. An optimized value can be selected during the readout board validation and later on during the commissioning phase. The time resolution has been measured with independent approaches, achieving results better than 150 ps, with the possibility to improve it further correcting for non-linearities in the TDC ramp. 
Whereas this correction is not needed for experiments using PMTs (which have a time resolution of few ns by their own), it offers an opportunity for the use of CATIROC beyond  PMT detectors (e.g. SiPM) and which require a resolution better than 100~ps. 
CATIROC offers some additional flexibility which may be of convenient to adapt to the specific requirements of different experiments. The configurable preamplifiers for each of the 16 input channels allow to compensate for differences in the gain of the PMT. 
This is particularly important when dealing with a very large number of PMTs, when a strict PMT selection is difficult to be handled.

Finally, we presented extensive tests with some 3-inch PMTs in order to compare the CATIROC response to the one obtained with an independent readout system. 
Further measurements, done with a large sample of JUNO PMTs and front-end boards confirmed these results but will be part of a future paper. 

\section*{Acknowledgments}
The authors thank all the JUNO-SPMT collaborators, Jacques Wurtz and Andrea Triossi for support and many fruitful discussions. %


\begin{thebibliography}{99}
\bibitem{JUNO-physics}JUNO collaboration, \emph{Neutrino physics with JUNO}, arXiv:1507.05613
\bibitem{JUNO-concept}JUNO collaboration, \emph{JUNO conceptual design report}, arXiv:1508.07166.
\bibitem{HyperK}Hyper-Kamiokande Proto-Collaboration, \emph{Hyper-Kamiokande Design Report}, arXiv:1805.04163. 
\bibitem{orca}KM3NeT Collaboration,  \emph{KM3NeT 2.0 : Letter of Intent for ARCA and ORCA }, J. Phys. G: Nucl. Part. Phys. 43 (2016) 084001. 
\bibitem{catiroc}S. Blin et. al. \emph{Performance of CATIROC: ASIC for smart readout of large photomultiplier arrays}, JINST 12 C03041.
\bibitem{parisroc}S. Conforti Di Lorenzo et al., \emph{PARISROC, an autonomous front-end ASIC for trigger-less acquisition in next generation neutrino experiments}, Nucl. Instum. Meth. A 695 (2012) 373.
\bibitem{PMm2} B. Genolini et al., \emph{PMm2: large photomultipliers and innovative electronics for the next-generation neutrino experiments}, Nucl. Instrum. Meth. A 610 (2009) 249, arXiv:0811.2681.
\bibitem{spmtMiao}Miao He, \emph{Double calorimetry system in JUNO}, Radiation Detection Technology and Methods, 2017, 1 (2) : 21.
\bibitem{SPMT-paper}Jilei Xu for the JUNO collaboration, \emph{The JUNO double calorimetry system}, Proceeding at NuFact 2019, 26 - August 31, 2019, Daegu, Korea.
\bibitem{bellamy}E.H. Bellamy, {\it et al.}, \emph{Absolute calibration and monitoring of a spectrometric channel using a photomultiplier}, Nucl. Instrum. Meth.  A 339 (1994) 468. 
\bibitem{hzc} Hainan Zhanchuang Photonics Technology Co., XP72B20 datasheet,  \href{http://www.hzcphotonics.com/products/XP72B20.pdf}{http://www.hzcphotonics.com/products/XP72B20.pdf.}
\end{thebibliography}
\end{document}